\documentclass[a4paper,traditabstract,longauth]{aa}  
\usepackage{graphicx}
\usepackage{lscape}
\usepackage{txfonts}
\usepackage[breaklinks, colorlinks, citecolor=blue]{hyperref}
\usepackage{url}
\usepackage{natbib}
\usepackage{color}
\usepackage{amssymb}
\usepackage{booktabs}
\usepackage{subfig}

\def\msol{{M$_{\odot}$}}
\def\xmm{XMM-{\it Newton} }

\def\xmm{{\it XMM-Newton}}
\def\chandra{{\it Chandra}}
\def\planck{{\it Planck}}
\def\rosat{{\it ROSAT}}
\def\rass{{\rm RASS}}

\def \mos {\hbox{\sc EMOS}} 
\def \pn {\hbox{\sc EPN}} 
\def \mekal {\hbox{\sc mekal}}

\newfont{\gwpfont}{cmssq8 scaled 1000}
\newcommand{\rexcess}{{\gwpfont REXCESS}}

\def\Mv{M_{500}}
\def\Rv{R_{500}}
\def\Mgv{M_{\rm g,500}}

\def\YX {Y_{\rm X}}
\def\TX {T_{\rm X}}

\def\YSZ {Y_{\rm SZ}}
\def\YSZ {Y_{500}}

\def\Mv {M_{\rm 500}}
\def \Rv {R_{500}}
\def\keV {\rm keV}

\def\MYX {$M_{500}$--$Y_{\rm X}$}

\def\toto{PLCK\,G266.6$-$27.3}

\def\msol {{\rm M_{\odot}}}

\def\lesssim{\mathrel{\hbox{\rlap{\hbox{\lower4pt\hbox{$\sim$}}}\hbox{$<$}}}}
\def\gtrsim{\mathrel{\hbox{\rlap{\hbox{\lower4pt\hbox{$\sim$}}}\hbox{$>$}}}}

\newcommand{\propsim}{\lower 3pt \hbox{$\, \buildrel {\textstyle
       \propto}\over {\textstyle \sim}\,$}}

\begin{document}
%

\author{\small
Planck Collaboration:
N.~Aghanim\inst{50}
\and
M.~Arnaud\inst{64}\thanks{Corresponding author: M. Arnaud,  monique.arnaud@cea.fr}
\and
M.~Ashdown\inst{62, 4}
\and
F.~Atrio-Barandela\inst{14}
\and
J.~Aumont\inst{50}
\and
C.~Baccigalupi\inst{73}
\and
A.~Balbi\inst{31}
\and
A.~J.~Banday\inst{82, 7, 68}
\and
R.~B.~Barreiro\inst{58}
% ...
}
\institute{\small
Aalto University Mets\"{a}hovi Radio Observatory, Mets\"{a}hovintie 114, FIN-02540 Kylm\"{a}l\"{a}, Finland\\
\and
Agenzia Spaziale Italiana Science Data Center, c/o ESRIN, via Galileo Galilei, Frascati, Italy\\
\and
Astroparticule et Cosmologie, CNRS (UMR7164), Universit\'{e} Denis Diderot Paris 7, B\^{a}timent Condorcet, 10 rue A. Domon et L\'{e}onie Duquet, Paris, France\\
\and
Astrophysics Group, Cavendish Laboratory, University of Cambridge, J J Thomson Avenue, Cambridge CB3 0HE, U.K.\\
\and
Atacama Large Millimeter/submillimeter Array, ALMA Santiago Central Offices, Alonso de Cordova 3107, Vitacura, Casilla 763 0355, Santiago, Chile\\
\and
CITA, University of Toronto, 60 St. George St., Toronto, ON M5S 3H8, Canada\\
\and
CNRS, IRAP, 9 Av. colonel Roche, BP 44346, F-31028 Toulouse cedex 4, France\\
\and
California Institute of Technology, Pasadena, California, U.S.A.\\
\and
Centre of Mathematics for Applications, University of Oslo, Blindern, Oslo, Norway\\
\and
Centro de Astrof\'{\i}sica, Universidade do Porto, Rua das Estrelas, 4150-762 Porto, Portugal\\
\and
Centro de Estudios de F\'{i}sica del Cosmos de Arag\'{o}n (CEFCA), Plaza San Juan, 1, planta 2, E-44001, Teruel, Spain\\
\and
DSM/Irfu/SPP, CEA-Saclay, F-91191 Gif-sur-Yvette Cedex, France\\
\and
DTU Space, National Space Institute, Juliane Mariesvej 30, Copenhagen, Denmark\\
\and
Departamento de F\'{\i}sica Fundamental, Facultad de Ciencias, Universidad de Salamanca, 37008 Salamanca, Spain\\
\and
Departamento de F\'{\i}sica, Universidad de Oviedo, Avda. Calvo Sotelo s/n, Oviedo, Spain\\
\and
Department of Astronomy and Astrophysics, University of Toronto, 50 Saint George Street, Toronto, Ontario, Canada\\
\and
Department of Physics \& Astronomy, University of British Columbia, 6224 Agricultural Road, Vancouver, British Columbia, Canada\\
\and
Department of Physics and Astronomy, University of Southern California, Los Angeles, California, U.S.A.\\
\and
Department of Physics and Astronomy, University of Sussex, Brighton BN1 9QH, U.K.\\
\and
Department of Physics, Gustaf H\"{a}llstr\"{o}min katu 2a, University of Helsinki, Helsinki, Finland\\
\and
Department of Physics, Princeton University, Princeton, New Jersey, U.S.A.\\
\and
Department of Physics, Purdue University, 525 Northwestern Avenue, West Lafayette, Indiana, U.S.A.\\
\and
Department of Physics, University of California, Berkeley, California, U.S.A.\\
\and
Department of Physics, University of California, Santa Barbara, California, U.S.A.\\
\and
Department of Physics, University of Illinois at Urbana-Champaign, 1110 West Green Street, Urbana, Illinois, U.S.A.\\
\and
Dipartimento di Fisica G. Galilei, Universit\`{a} degli Studi di Padova, via Marzolo 8, 35131 Padova, Italy\\
\and
Dipartimento di Fisica, Universit\`{a} La Sapienza, P. le A. Moro 2, Roma, Italy\\
\and
Dipartimento di Fisica, Universit\`{a} degli Studi di Milano, Via Celoria, 16, Milano, Italy\\
\and
Dipartimento di Fisica, Universit\`{a} degli Studi di Trieste, via A. Valerio 2, Trieste, Italy\\
\and
Dipartimento di Fisica, Universit\`{a} di Ferrara, Via Saragat 1, 44122 Ferrara, Italy\\
\and
Dipartimento di Fisica, Universit\`{a} di Roma Tor Vergata, Via della Ricerca Scientifica, 1, Roma, Italy\\
\and
Discovery Center, Niels Bohr Institute, Blegdamsvej 17, Copenhagen, Denmark\\
\and
Dpto. Astrof\'{i}sica, Universidad de La Laguna (ULL), E-38206 La Laguna, Tenerife, Spain\\
\and
European Southern Observatory, ESO Vitacura, Alonso de Cordova 3107, Vitacura, Casilla 19001, Santiago, Chile\\
\and
European Space Agency, ESAC, Camino bajo del Castillo, s/n, Urbanizaci\'{o}n Villafranca del Castillo, Villanueva de la Ca\~{n}ada, Madrid, Spain\\
\and
European Space Agency, ESAC, Planck Science Office, Camino bajo del Castillo, s/n, Urbanizaci\'{o}n Villafranca del Castillo, Villanueva de la Ca\~{n}ada, Madrid, Spain\\
\and
European Space Agency, ESTEC, Keplerlaan 1, 2201 AZ Noordwijk, The Netherlands\\
\and
Finnish Centre for Astronomy with ESO (FINCA), University of Turku, V\"{a}is\"{a}l\"{a}ntie 20, FIN-21500, Piikki\"{o}, Finland\\
\and
Helsinki Institute of Physics, Gustaf H\"{a}llstr\"{o}min katu 2, University of Helsinki, Helsinki, Finland\\
\and
INAF - Osservatorio Astronomico di Padova, Vicolo dell'Osservatorio 5, Padova, Italy\\
\and
INAF - Osservatorio Astronomico di Roma, via di Frascati 33, Monte Porzio Catone, Italy\\
\and
INAF - Osservatorio Astronomico di Trieste, Via G.B. Tiepolo 11, Trieste, Italy\\
\and
INAF/IASF Bologna, Via Gobetti 101, Bologna, Italy\\
\and
INAF/IASF Milano, Via E. Bassini 15, Milano, Italy\\
\and
IPAG: Institut de Plan\'{e}tologie et d'Astrophysique de Grenoble, Universit\'{e} Joseph Fourier, Grenoble 1 / CNRS-INSU, UMR 5274, Grenoble, F-38041, France\\
\and
ISDC Data Centre for Astrophysics, University of Geneva, ch. d'Ecogia 16, Versoix, Switzerland\\
\and
Imperial College London, Astrophysics group, Blackett Laboratory, Prince Consort Road, London, SW7 2AZ, U.K.\\
\and
Infrared Processing and Analysis Center, California Institute of Technology, Pasadena, CA 91125, U.S.A.\\
\and
Institut N\'{e}el, CNRS, Universit\'{e} Joseph Fourier Grenoble I, 25 rue des Martyrs, Grenoble, France\\
\and
Institut d'Astrophysique Spatiale, CNRS (UMR8617) Universit\'{e} Paris-Sud 11, B\^{a}timent 121, Orsay, France\\
\and
Institut d'Astrophysique de Paris, CNRS (UMR7095), 98 bis Boulevard Arago, F-75014, Paris, France\\
\and
Institut de Ci\`{e}ncies de l'Espai, CSIC/IEEC, Facultat de Ci\`{e}ncies, Campus UAB, Torre C5 par-2, Bellaterra 08193, Spain\\
\and
Institute for Space Sciences, Bucharest-Magurale, Romania\\
\and
Institute of Astronomy and Astrophysics, Academia Sinica, Taipei, Taiwan\\
\and
Institute of Astronomy, University of Cambridge, Madingley Road, Cambridge CB3 0HA, U.K.\\
\and
Institute of Theoretical Astrophysics, University of Oslo, Blindern, Oslo, Norway\\
\and
Instituto de Astrof\'{\i}sica de Canarias, C/V\'{\i}a L\'{a}ctea s/n, La Laguna, Tenerife, Spain\\
\and
Instituto de F\'{\i}sica de Cantabria (CSIC-Universidad de Cantabria), Avda. de los Castros s/n, Santander, Spain\\
\and
Istituto di Fisica del Plasma, CNR-ENEA-EURATOM Association, Via R. Cozzi 53, Milano, Italy\\
\and
Jet Propulsion Laboratory, California Institute of Technology, 4800 Oak Grove Drive, Pasadena, California, U.S.A.\\
\and
Jodrell Bank Centre for Astrophysics, Alan Turing Building, School of Physics and Astronomy, The University of Manchester, Oxford Road, Manchester, M13 9PL, U.K.\\
\and
Kavli Institute for Cosmology Cambridge, Madingley Road, Cambridge, CB3 0HA, U.K.\\
\and
LERMA, CNRS, Observatoire de Paris, 61 Avenue de l'Observatoire, Paris, France\\
\and
Laboratoire AIM, IRFU/Service d'Astrophysique - CEA/DSM - CNRS - Universit\'{e} Paris Diderot, B\^{a}t. 709, CEA-Saclay, F-91191 Gif-sur-Yvette Cedex, France\\
\and
Laboratoire de Physique Subatomique et de Cosmologie, CNRS/IN2P3, Universit\'{e} Joseph Fourier Grenoble I, Institut National Polytechnique de Grenoble, 53 rue des Martyrs, 38026 Grenoble cedex, France\\
\and
Laboratoire de l'Acc\'{e}l\'{e}rateur Lin\'{e}aire, Universit\'{e} Paris-Sud 11, CNRS/IN2P3, Orsay, France\\
\and
Lawrence Berkeley National Laboratory, Berkeley, California, U.S.A.\\
\and
Max-Planck-Institut f\"{u}r Astrophysik, Karl-Schwarzschild-Str. 1, 85741 Garching, Germany\\
\and
Max-Planck-Institut f\"{u}r Extraterrestrische Physik, Giessenbachstra{\ss}e, 85748 Garching, Germany\\
\and
Niels Bohr Institute, Blegdamsvej 17, Copenhagen, Denmark\\
\and
Observational Cosmology, Mail Stop 367-17, California Institute of Technology, Pasadena, CA, 91125, U.S.A.\\
\and
Optical Science Laboratory, University College London, Gower Street, London, U.K.\\
\and
SISSA, Astrophysics Sector, via Bonomea 265, 34136, Trieste, Italy\\
\and
School of Physics and Astronomy, Cardiff University, Queens Buildings, The Parade, Cardiff, CF24 3AA, U.K.\\
\and
Space Research Institute (IKI), Russian Academy of Sciences, Profsoyuznaya Str, 84/32, Moscow, 117997, Russia\\
\and
Space Sciences Laboratory, University of California, Berkeley, California, U.S.A.\\
\and
Stanford University, Dept of Physics, Varian Physics Bldg, 382 Via Pueblo Mall, Stanford, California, U.S.A.\\
\and
Tartu Observatory, Toravere, Tartumaa, 61602, Estonia\\
\and
Tuorla Observatory, Department of Physics and Astronomy, University of Turku, V\"ais\"al\"antie 20, FIN-21500, Piikki\"o, Finland\\
\and
UPMC Univ Paris 06, UMR7095, 98 bis Boulevard Arago, F-75014, Paris, France\\
\and
Universit\"{a}t Heidelberg, Institut f\"{u}r Theoretische Astrophysik, Albert-\"{U}berle-Str. 2, 69120, Heidelberg, Germany\\
\and
Universit\'{e} de Toulouse, UPS-OMP, IRAP, F-31028 Toulouse cedex 4, France\\
% ...
}

\title{\textit{Planck} Early Results XXVI:  Detection with \textit{Planck}  and confirmation by \textit{XMM-Newton} of PLCK~G266.6$-$27.3,  an exceptionally X-ray luminous and massive galaxy cluster at $z\sim1$}
 \date{Received June 7; Accepted July 11}
  \abstract
  {We present first results on PLCK\,G266.6$-$27.3, a galaxy cluster candidate detected at a signal-to-noise ratio of $5$ in the \planck\ All Sky survey. An \xmm\ validation observation has allowed us to confirm that the candidate is a {\it bona fide} galaxy cluster. With these X-ray data we measure an accurate redshift, $z = 0.94 \pm 0.02$, and estimate the cluster mass to be $\Mv = (7.8 \pm0.8)\times10^{14}\,\msol$. PLCK\,G266.6$-$27.3 is an exceptional system: its luminosity of $L_{\rm X}[0.5$--$2.0\keV]=(1.4\pm0.05)\times 10^{45}\,{\rm erg\,s^{-1}}$  equals that of the two most luminous known clusters in the $z > 0.5$ universe, and it is one of the most massive clusters at $z\sim1$. Moreover, unlike the majority of high-redshift clusters, PLCK\,G266.6$-$27.3 appears to be highly relaxed. This observation confirms \planck's capability of detecting high-redshift, high-mass clusters, and opens the way to the systematic study of population evolution in the exponential tail of the mass function.}
  

\authorrunning{Planck Collaboration}
\titlerunning{PLCK~G266.6-27.3,  an exceptionally X-ray luminous galaxy cluster at $z\sim1$}
  \maketitle

\def\setsymbol#1#2{\expandafter\def\csname #1\endcsname{#2}}
\def\getsymbol#1{\csname #1\endcsname}

\def\Planck{{\it Planck\/}}

\def\HeJT{$^4$He-JT}

\def\allearlypapers{\nocite{planck2011-1.1, planck2011-1.3, planck2011-1.4, planck2011-1.5, planck2011-1.6, planck2011-1.7, planck2011-1.10, planck2011-1.10sup, planck2011-5.1a, planck2011-5.1b, planck2011-5.2a, planck2011-5.2b, planck2011-5.2c, planck2011-6.1, planck2011-6.2, planck2011-6.3a, planck2011-6.4a, planck2011-6.4b, planck2011-6.6, planck2011-7.0, planck2011-7.2, planck2011-7.3, planck2011-7.7a, planck2011-7.7b, planck2011-7.12, planck2011-7.13}}

\newbox\tablebox    \newdimen\tablewidth
\def\leaderfil{\leaders\hbox to 5pt{\hss.\hss}\hfil}
%
%
\def\endPlancktable{\tablewidth=\columnwidth 
    $$\hss\copy\tablebox\hss$$
    \vskip-\lastskip\vskip -2pt}
\def\endPlancktablewide{\tablewidth=\textwidth 
    $$\hss\copy\tablebox\hss$$
    \vskip-\lastskip\vskip -2pt}
\def\tablenote#1 #2\par{\begingroup \parindent=0.8em
    \abovedisplayshortskip=0pt\belowdisplayshortskip=0pt
    \noindent
    $$\hss\vbox{\hsize\tablewidth \hangindent=\parindent \hangafter=1 \noindent
    \hbox to \parindent{\sup{\rm #1}\hss}\strut#2\strut\par}\hss$$
    \endgroup}
\def\doubleline{\vskip 3pt\hrule \vskip 1.5pt \hrule \vskip 5pt}

%
\def\L2{\ifmmode L_2\else $L_2$\fi}
\def\dtt{\Delta T/T}
\def\DeltaT{\ifmmode \Delta T\else $\Delta T$\fi}
\def\deltat{\ifmmode \Delta t\else $\Delta t$\fi}
\def\fknee{\ifmmode f_{\rm knee}\else $f_{\rm knee}$\fi}
\def\Fmax{\ifmmode F_{\rm max}\else $F_{\rm max}$\fi}
\def\solar{\ifmmode{\rm M}_{\mathord\odot}\else${\rm M}_{\mathord\odot}$\fi}
\def\sol{\solar}
\def\mag{\sup{m}}
\def\inv{\ifmmode^{-1}\else$^{-1}$\fi}
\def\mo{\ifmmode^{-1}\else$^{-1}$\fi}
\def\sup#1{\ifmmode ^{\rm #1}\else $^{\rm #1}$\fi}
\def\expo#1{\ifmmode \times 10^{#1}\else $\times 10^{#1}$\fi}
\def\,{\thinspace}
\def\lsim{\mathrel{\raise .4ex\hbox{\rlap{$<$}\lower 1.2ex\hbox{$\sim$}}}}
\def\gsim{\mathrel{\raise .4ex\hbox{\rlap{$>$}\lower 1.2ex\hbox{$\sim$}}}}
\let\lea=\lsim
\let\gea=\gsim
\def\simprop{\mathrel{\raise .4ex\hbox{\rlap{$\propto$}\lower 1.2ex\hbox{$\sim$}}}}
\def\deg{\ifmmode^\circ\else$^\circ$\fi}
\def\pdeg{\ifmmode $\setbox0=\hbox{$^{\circ}$}\rlap{\hskip.11\wd0 .}$^{\circ}
          \else \setbox0=\hbox{$^{\circ}$}\rlap{\hskip.11\wd0 .}$^{\circ}$\fi}
\def\arcs{\ifmmode {^{\scriptstyle\prime\prime}}
          \else $^{\scriptstyle\prime\prime}$\fi}
\def\arcm{\ifmmode {^{\scriptstyle\prime}}
          \else $^{\scriptstyle\prime}$\fi}
\newdimen\sa  \newdimen\sb
\def\parcs{\sa=.07em \sb=.03em
     \ifmmode \hbox{\rlap{.}}^{\scriptstyle\prime\kern -\sb\prime}\hbox{\kern -\sa}
     \else \rlap{.}$^{\scriptstyle\prime\kern -\sb\prime}$\kern -\sa\fi}
\def\parcm{\sa=.08em \sb=.03em
     \ifmmode \hbox{\rlap{.}\kern\sa}^{\scriptstyle\prime}\hbox{\kern-\sb}
     \else \rlap{.}\kern\sa$^{\scriptstyle\prime}$\kern-\sb\fi}
\def\ra[#1 #2 #3.#4]{#1\sup{h}#2\sup{m}#3\sup{s}\llap.#4}
\def\dec[#1 #2 #3.#4]{#1\deg#2\arcm#3\arcs\llap.#4}
\def\deco[#1 #2 #3]{#1\deg#2\arcm#3\arcs}
\def\rra[#1 #2]{#1\sup{h}#2\sup{m}}
\def\page{\vfill\eject}
\def\dots{\relax\ifmmode \ldots\else $\ldots$\fi}
%
%
\def\WHzsr{\ifmmode $W\,Hz\mo\,sr\mo$\else W\,Hz\mo\,sr\mo\fi}
\def\mHz{\ifmmode $\,mHz$\else \,mHz\fi}
\def\GHz{\ifmmode $\,GHz$\else \,GHz\fi}
\def\mKs{\ifmmode $\,mK\,s$^{1/2}\else \,mK\,s$^{1/2}$\fi}
\def\muKs{\ifmmode \,\mu$K\,s$^{1/2}\else \,$\mu$K\,s$^{1/2}$\fi}
\def\muKRJs{\ifmmode \,\mu$K$_{\rm RJ}$\,s$^{1/2}\else \,$\mu$K$_{\rm RJ}$\,s$^{1/2}$\fi}
\def\muKHz{\ifmmode \,\mu$K\,Hz$^{-1/2}\else \,$\mu$K\,Hz$^{-1/2}$\fi}
\def\MJysr{\ifmmode \,$MJy\,sr\mo$\else \,MJy\,sr\mo\fi}
\def\MJysrmK{\ifmmode \,$MJy\,sr\mo$\,mK$_{\rm CMB}\mo\else \,MJy\,sr\mo\,mK$_{\rm CMB}\mo$\fi}
\def\microns{\ifmmode \,\mu$m$\else \,$\mu$m\fi}
\def\micron{\microns}
\def\muK{\ifmmode \,\mu$K$\else \,$\mu$\hbox{K}\fi}
\def\microK{\ifmmode \,\mu$K$\else \,$\mu$\hbox{K}\fi}
\def\muW{\ifmmode \,\mu$W$\else \,$\mu$\hbox{W}\fi}
\def\kms{\ifmmode $\,km\,s$^{-1}\else \,km\,s$^{-1}$\fi}
\def\kmsMpc{\ifmmode $\,\kms\,Mpc\mo$\else \,\kms\,Mpc\mo\fi}
%
%


\setsymbol{LFI:center:frequency:70GHz:units}{70.3\,GHz}
\setsymbol{LFI:center:frequency:44GHz:units}{44.1\,GHz}
\setsymbol{LFI:center:frequency:30GHz:units}{28.5\,GHz}

\setsymbol{LFI:center:frequency:70GHz}{70.3}
\setsymbol{LFI:center:frequency:44GHz}{44.1}
\setsymbol{LFI:center:frequency:30GHz}{28.5}

\setsymbol{LFI:center:frequency:LFI18:Rad:M:units}{71.7\GHz}
\setsymbol{LFI:center:frequency:LFI19:Rad:M:units}{67.5\GHz}
\setsymbol{LFI:center:frequency:LFI20:Rad:M:units}{69.2\GHz}
\setsymbol{LFI:center:frequency:LFI21:Rad:M:units}{70.4\GHz}
\setsymbol{LFI:center:frequency:LFI22:Rad:M:units}{71.5\GHz}
\setsymbol{LFI:center:frequency:LFI23:Rad:M:units}{70.8\GHz}
\setsymbol{LFI:center:frequency:LFI24:Rad:M:units}{44.4\GHz}
\setsymbol{LFI:center:frequency:LFI25:Rad:M:units}{44.0\GHz}
\setsymbol{LFI:center:frequency:LFI26:Rad:M:units}{43.9\GHz}
\setsymbol{LFI:center:frequency:LFI27:Rad:M:units}{28.3\GHz}
\setsymbol{LFI:center:frequency:LFI28:Rad:M:units}{28.8\GHz}
\setsymbol{LFI:center:frequency:LFI18:Rad:S:units}{70.1\GHz}
\setsymbol{LFI:center:frequency:LFI19:Rad:S:units}{69.6\GHz}
\setsymbol{LFI:center:frequency:LFI20:Rad:S:units}{69.5\GHz}
\setsymbol{LFI:center:frequency:LFI21:Rad:S:units}{69.5\GHz}
\setsymbol{LFI:center:frequency:LFI22:Rad:S:units}{72.8\GHz}
\setsymbol{LFI:center:frequency:LFI23:Rad:S:units}{71.3\GHz}
\setsymbol{LFI:center:frequency:LFI24:Rad:S:units}{44.1\GHz}
\setsymbol{LFI:center:frequency:LFI25:Rad:S:units}{44.1\GHz}
\setsymbol{LFI:center:frequency:LFI26:Rad:S:units}{44.1\GHz}
\setsymbol{LFI:center:frequency:LFI27:Rad:S:units}{28.5\GHz}
\setsymbol{LFI:center:frequency:LFI28:Rad:S:units}{28.2\GHz}

\setsymbol{LFI:center:frequency:LFI18:Rad:M}{71.7}
\setsymbol{LFI:center:frequency:LFI19:Rad:M}{67.5}
\setsymbol{LFI:center:frequency:LFI20:Rad:M}{69.2}
\setsymbol{LFI:center:frequency:LFI21:Rad:M}{70.4}
\setsymbol{LFI:center:frequency:LFI22:Rad:M}{71.5}
\setsymbol{LFI:center:frequency:LFI23:Rad:M}{70.8}
\setsymbol{LFI:center:frequency:LFI24:Rad:M}{44.4}
\setsymbol{LFI:center:frequency:LFI25:Rad:M}{44.0}
\setsymbol{LFI:center:frequency:LFI26:Rad:M}{43.9}
\setsymbol{LFI:center:frequency:LFI27:Rad:M}{28.3}
\setsymbol{LFI:center:frequency:LFI28:Rad:M}{28.8}
\setsymbol{LFI:center:frequency:LFI18:Rad:S}{70.1}
\setsymbol{LFI:center:frequency:LFI19:Rad:S}{69.6}
\setsymbol{LFI:center:frequency:LFI20:Rad:S}{69.5}
\setsymbol{LFI:center:frequency:LFI21:Rad:S}{69.5}
\setsymbol{LFI:center:frequency:LFI22:Rad:S}{72.8}
\setsymbol{LFI:center:frequency:LFI23:Rad:S}{71.3}
\setsymbol{LFI:center:frequency:LFI24:Rad:S}{44.1}
\setsymbol{LFI:center:frequency:LFI25:Rad:S}{44.1}
\setsymbol{LFI:center:frequency:LFI26:Rad:S}{44.1}
\setsymbol{LFI:center:frequency:LFI27:Rad:S}{28.5}
\setsymbol{LFI:center:frequency:LFI28:Rad:S}{28.2}


\setsymbol{LFI:white:noise:sensitivity:70GHz:units}{152.6\muKs}
\setsymbol{LFI:white:noise:sensitivity:44GHz:units}{173.1\muKs}
\setsymbol{LFI:white:noise:sensitivity:30GHz:units}{146.8\muKs}

\setsymbol{LFI:white:noise:sensitivity:70GHz}{152.6}
\setsymbol{LFI:white:noise:sensitivity:44GHz}{173.1}
\setsymbol{LFI:white:noise:sensitivity:30GHz}{146.8}

\setsymbol{LFI:white:noise:sensitivity:LFI18:Rad:M:units}{512.0\muKs}
\setsymbol{LFI:white:noise:sensitivity:LFI19:Rad:M:units}{581.4\muKs}
\setsymbol{LFI:white:noise:sensitivity:LFI20:Rad:M:units}{590.8\muKs}
\setsymbol{LFI:white:noise:sensitivity:LFI21:Rad:M:units}{455.2\muKs}
\setsymbol{LFI:white:noise:sensitivity:LFI22:Rad:M:units}{492.0\muKs}
\setsymbol{LFI:white:noise:sensitivity:LFI23:Rad:M:units}{507.7\muKs}
\setsymbol{LFI:white:noise:sensitivity:LFI24:Rad:M:units}{462.2\muKs}
\setsymbol{LFI:white:noise:sensitivity:LFI25:Rad:M:units}{413.6\muKs}
\setsymbol{LFI:white:noise:sensitivity:LFI26:Rad:M:units}{478.6\muKs}
\setsymbol{LFI:white:noise:sensitivity:LFI27:Rad:M:units}{277.7\muKs}
\setsymbol{LFI:white:noise:sensitivity:LFI28:Rad:M:units}{312.3\muKs}
\setsymbol{LFI:white:noise:sensitivity:LFI18:Rad:S:units}{465.7\muKs}
\setsymbol{LFI:white:noise:sensitivity:LFI19:Rad:S:units}{555.6\muKs}
\setsymbol{LFI:white:noise:sensitivity:LFI20:Rad:S:units}{623.2\muKs}
\setsymbol{LFI:white:noise:sensitivity:LFI21:Rad:S:units}{564.1\muKs}
\setsymbol{LFI:white:noise:sensitivity:LFI22:Rad:S:units}{534.4\muKs}
\setsymbol{LFI:white:noise:sensitivity:LFI23:Rad:S:units}{542.4\muKs}
\setsymbol{LFI:white:noise:sensitivity:LFI24:Rad:S:units}{399.2\muKs}
\setsymbol{LFI:white:noise:sensitivity:LFI25:Rad:S:units}{392.6\muKs}
\setsymbol{LFI:white:noise:sensitivity:LFI26:Rad:S:units}{418.6\muKs}
\setsymbol{LFI:white:noise:sensitivity:LFI27:Rad:S:units}{302.9\muKs}
\setsymbol{LFI:white:noise:sensitivity:LFI28:Rad:S:units}{285.3\muKs}

\setsymbol{LFI:white:noise:sensitivity:LFI18:Rad:M}{512.0}
\setsymbol{LFI:white:noise:sensitivity:LFI19:Rad:M}{581.4}
\setsymbol{LFI:white:noise:sensitivity:LFI20:Rad:M}{590.8}
\setsymbol{LFI:white:noise:sensitivity:LFI21:Rad:M}{455.2}
\setsymbol{LFI:white:noise:sensitivity:LFI22:Rad:M}{492.0}
\setsymbol{LFI:white:noise:sensitivity:LFI23:Rad:M}{507.7}
\setsymbol{LFI:white:noise:sensitivity:LFI24:Rad:M}{462.2}
\setsymbol{LFI:white:noise:sensitivity:LFI25:Rad:M}{413.6}
\setsymbol{LFI:white:noise:sensitivity:LFI26:Rad:M}{478.6}
\setsymbol{LFI:white:noise:sensitivity:LFI27:Rad:M}{277.7}
\setsymbol{LFI:white:noise:sensitivity:LFI28:Rad:M}{312.3}
\setsymbol{LFI:white:noise:sensitivity:LFI18:Rad:S}{465.7}
\setsymbol{LFI:white:noise:sensitivity:LFI19:Rad:S}{555.6}
\setsymbol{LFI:white:noise:sensitivity:LFI20:Rad:S}{623.2}
\setsymbol{LFI:white:noise:sensitivity:LFI21:Rad:S}{564.1}
\setsymbol{LFI:white:noise:sensitivity:LFI22:Rad:S}{534.4}
\setsymbol{LFI:white:noise:sensitivity:LFI23:Rad:S}{542.4}
\setsymbol{LFI:white:noise:sensitivity:LFI24:Rad:S}{399.2}
\setsymbol{LFI:white:noise:sensitivity:LFI25:Rad:S}{392.6}
\setsymbol{LFI:white:noise:sensitivity:LFI26:Rad:S}{418.6}
\setsymbol{LFI:white:noise:sensitivity:LFI27:Rad:S}{302.9}
\setsymbol{LFI:white:noise:sensitivity:LFI28:Rad:S}{285.3}


\setsymbol{LFI:knee:frequency:70GHz:units}{29.5\mHz}
\setsymbol{LFI:knee:frequency:44GHz:units}{56.2\mHz}
\setsymbol{LFI:knee:frequency:30GHz:units}{113.7\mHz}

\setsymbol{LFI:knee:frequency:70GHz}{29.5}
\setsymbol{LFI:knee:frequency:44GHz}{56.2}
\setsymbol{LFI:knee:frequency:30GHz}{113.7}

\setsymbol{LFI:knee:frequency:LFI18:Rad:M:units}{16.3\mHz}
\setsymbol{LFI:knee:frequency:LFI19:Rad:M:units}{15.1\mHz}
\setsymbol{LFI:knee:frequency:LFI20:Rad:M:units}{18.7\mHz}
\setsymbol{LFI:knee:frequency:LFI21:Rad:M:units}{37.2\mHz}
\setsymbol{LFI:knee:frequency:LFI22:Rad:M:units}{12.7\mHz}
\setsymbol{LFI:knee:frequency:LFI23:Rad:M:units}{34.6\mHz}
\setsymbol{LFI:knee:frequency:LFI24:Rad:M:units}{46.2\mHz}
\setsymbol{LFI:knee:frequency:LFI25:Rad:M:units}{24.9\mHz}
\setsymbol{LFI:knee:frequency:LFI26:Rad:M:units}{67.6\mHz}
\setsymbol{LFI:knee:frequency:LFI27:Rad:M:units}{187.4\mHz}
\setsymbol{LFI:knee:frequency:LFI28:Rad:M:units}{122.2\mHz}
\setsymbol{LFI:knee:frequency:LFI18:Rad:S:units}{17.7\mHz}
\setsymbol{LFI:knee:frequency:LFI19:Rad:S:units}{22.0\mHz}
\setsymbol{LFI:knee:frequency:LFI20:Rad:S:units}{8.7\mHz}
\setsymbol{LFI:knee:frequency:LFI21:Rad:S:units}{25.9\mHz}
\setsymbol{LFI:knee:frequency:LFI22:Rad:S:units}{15.8\mHz}
\setsymbol{LFI:knee:frequency:LFI23:Rad:S:units}{129.8\mHz}
\setsymbol{LFI:knee:frequency:LFI24:Rad:S:units}{100.9\mHz}
\setsymbol{LFI:knee:frequency:LFI25:Rad:S:units}{38.9\mHz}
\setsymbol{LFI:knee:frequency:LFI26:Rad:S:units}{58.9\mHz}
\setsymbol{LFI:knee:frequency:LFI27:Rad:S:units}{104.4\mHz}
\setsymbol{LFI:knee:frequency:LFI28:Rad:S:units}{40.7\mHz}

\setsymbol{LFI:knee:frequency:LFI18:Rad:M}{16.3}
\setsymbol{LFI:knee:frequency:LFI19:Rad:M}{15.1}
\setsymbol{LFI:knee:frequency:LFI20:Rad:M}{18.7}
\setsymbol{LFI:knee:frequency:LFI21:Rad:M}{37.2}
\setsymbol{LFI:knee:frequency:LFI22:Rad:M}{12.7}
\setsymbol{LFI:knee:frequency:LFI23:Rad:M}{34.6}
\setsymbol{LFI:knee:frequency:LFI24:Rad:M}{46.2}
\setsymbol{LFI:knee:frequency:LFI25:Rad:M}{24.9}
\setsymbol{LFI:knee:frequency:LFI26:Rad:M}{67.6}
\setsymbol{LFI:knee:frequency:LFI27:Rad:M}{187.4}
\setsymbol{LFI:knee:frequency:LFI28:Rad:M}{122.2}
\setsymbol{LFI:knee:frequency:LFI18:Rad:S}{17.7}
\setsymbol{LFI:knee:frequency:LFI19:Rad:S}{22.0}
\setsymbol{LFI:knee:frequency:LFI20:Rad:S}{8.7}
\setsymbol{LFI:knee:frequency:LFI21:Rad:S}{25.9}
\setsymbol{LFI:knee:frequency:LFI22:Rad:S}{15.8}
\setsymbol{LFI:knee:frequency:LFI23:Rad:S}{129.8}
\setsymbol{LFI:knee:frequency:LFI24:Rad:S}{100.9}
\setsymbol{LFI:knee:frequency:LFI25:Rad:S}{38.9}
\setsymbol{LFI:knee:frequency:LFI26:Rad:S}{58.9}
\setsymbol{LFI:knee:frequency:LFI27:Rad:S}{104.4}
\setsymbol{LFI:knee:frequency:LFI28:Rad:S}{40.7}


\setsymbol{LFI:slope:70GHz:units}{$-1.03$\mHz}
\setsymbol{LFI:slope:44GHz:units}{$-0.89$\mHz}
\setsymbol{LFI:slope:30GHz:units}{$-0.87$\mHz}

\setsymbol{LFI:slope:70GHz}{$-1.03$}
\setsymbol{LFI:slope:44GHz}{$-0.89$}
\setsymbol{LFI:slope:30GHz}{$-0.87$}

\setsymbol{LFI:slope:LFI18:Rad:M:units}{$-1.04$\mHz}
\setsymbol{LFI:slope:LFI19:Rad:M:units}{$-1.09$\mHz}
\setsymbol{LFI:slope:LFI20:Rad:M:units}{$-0.69$\mHz}
\setsymbol{LFI:slope:LFI21:Rad:M:units}{$-1.56$\mHz}
\setsymbol{LFI:slope:LFI22:Rad:M:units}{$-1.01$\mHz}
\setsymbol{LFI:slope:LFI23:Rad:M:units}{$-0.96$\mHz}
\setsymbol{LFI:slope:LFI24:Rad:M:units}{$-0.83$\mHz}
\setsymbol{LFI:slope:LFI25:Rad:M:units}{$-0.91$\mHz}
\setsymbol{LFI:slope:LFI26:Rad:M:units}{$-0.95$\mHz}
\setsymbol{LFI:slope:LFI27:Rad:M:units}{$-0.87$\mHz}
\setsymbol{LFI:slope:LFI28:Rad:M:units}{$-0.88$\mHz}
\setsymbol{LFI:slope:LFI18:Rad:S:units}{$-1.15$\mHz}
\setsymbol{LFI:slope:LFI19:Rad:S:units}{$-1.00$\mHz}
\setsymbol{LFI:slope:LFI20:Rad:S:units}{$-0.95$\mHz}
\setsymbol{LFI:slope:LFI21:Rad:S:units}{$-0.92$\mHz}
\setsymbol{LFI:slope:LFI22:Rad:S:units}{$-1.01$\mHz}
\setsymbol{LFI:slope:LFI23:Rad:S:units}{$-0.95$\mHz}
\setsymbol{LFI:slope:LFI24:Rad:S:units}{$-0.73$\mHz}
\setsymbol{LFI:slope:LFI25:Rad:S:units}{$-1.16$\mHz}
\setsymbol{LFI:slope:LFI26:Rad:S:units}{$-0.79$\mHz}
\setsymbol{LFI:slope:LFI27:Rad:S:units}{$-0.82$\mHz}
\setsymbol{LFI:slope:LFI28:Rad:S:units}{$-0.91$\mHz}

\setsymbol{LFI:slope:LFI18:Rad:M}{$-1.04$}
\setsymbol{LFI:slope:LFI19:Rad:M}{$-1.09$}
\setsymbol{LFI:slope:LFI20:Rad:M}{$-0.69$}
\setsymbol{LFI:slope:LFI21:Rad:M}{$-1.56$}
\setsymbol{LFI:slope:LFI22:Rad:M}{$-1.01$}
\setsymbol{LFI:slope:LFI23:Rad:M}{$-0.96$}
\setsymbol{LFI:slope:LFI24:Rad:M}{$-0.83$}
\setsymbol{LFI:slope:LFI25:Rad:M}{$-0.91$}
\setsymbol{LFI:slope:LFI26:Rad:M}{$-0.95$}
\setsymbol{LFI:slope:LFI27:Rad:M}{$-0.87$}
\setsymbol{LFI:slope:LFI28:Rad:M}{$-0.88$}
\setsymbol{LFI:slope:LFI18:Rad:S}{$-1.15$}
\setsymbol{LFI:slope:LFI19:Rad:S}{$-1.00$}
\setsymbol{LFI:slope:LFI20:Rad:S}{$-0.95$}
\setsymbol{LFI:slope:LFI21:Rad:S}{$-0.92$}
\setsymbol{LFI:slope:LFI22:Rad:S}{$-1.01$}
\setsymbol{LFI:slope:LFI23:Rad:S}{$-0.95$}
\setsymbol{LFI:slope:LFI24:Rad:S}{$-0.73$}
\setsymbol{LFI:slope:LFI25:Rad:S}{$-1.16$}
\setsymbol{LFI:slope:LFI26:Rad:S}{$-0.79$}
\setsymbol{LFI:slope:LFI27:Rad:S}{$-0.82$}
\setsymbol{LFI:slope:LFI28:Rad:S}{$-0.91$}


\setsymbol{LFI:FWHM:70GHz:units}{13\parcm01}
\setsymbol{LFI:FWHM:44GHz:units}{27\parcm92}
\setsymbol{LFI:FWHM:30GHz:units}{32\parcm65}

\setsymbol{LFI:FWHM:70GHz}{13.01}
\setsymbol{LFI:FWHM:44GHz}{27.92}
\setsymbol{LFI:FWHM:30GHz}{32.65}

\setsymbol{LFI:FWHM:LFI18:units}{13\parcm39}
\setsymbol{LFI:FWHM:LFI19:units}{13\parcm01}
\setsymbol{LFI:FWHM:LFI20:units}{12\parcm75}
\setsymbol{LFI:FWHM:LFI21:units}{12\parcm74}
\setsymbol{LFI:FWHM:LFI22:units}{12\parcm87}
\setsymbol{LFI:FWHM:LFI23:units}{13\parcm27}
\setsymbol{LFI:FWHM:LFI24:units}{22\parcm98}
\setsymbol{LFI:FWHM:LFI25:units}{30\parcm46}
\setsymbol{LFI:FWHM:LFI26:units}{30\parcm31}
\setsymbol{LFI:FWHM:LFI27:units}{32\parcm65}
\setsymbol{LFI:FWHM:LFI28:units}{32\parcm66}

\setsymbol{LFI:FWHM:LFI18}{13.39}
\setsymbol{LFI:FWHM:LFI19}{13.01}
\setsymbol{LFI:FWHM:LFI20}{12.75}
\setsymbol{LFI:FWHM:LFI21}{12.74}
\setsymbol{LFI:FWHM:LFI22}{12.87}
\setsymbol{LFI:FWHM:LFI23}{13.27}
\setsymbol{LFI:FWHM:LFI24}{22.98}
\setsymbol{LFI:FWHM:LFI25}{30.46}
\setsymbol{LFI:FWHM:LFI26}{30.31}
\setsymbol{LFI:FWHM:LFI27}{32.65}
\setsymbol{LFI:FWHM:LFI28}{32.66}



\setsymbol{LFI:FWHM:uncertainty:LFI18:units}{0.170\arcm}
\setsymbol{LFI:FWHM:uncertainty:LFI19:units}{0.174\arcm}
\setsymbol{LFI:FWHM:uncertainty:LFI20:units}{0.170\arcm}
\setsymbol{LFI:FWHM:uncertainty:LFI21:units}{0.156\arcm}
\setsymbol{LFI:FWHM:uncertainty:LFI22:units}{0.164\arcm}
\setsymbol{LFI:FWHM:uncertainty:LFI23:units}{0.171\arcm}
\setsymbol{LFI:FWHM:uncertainty:LFI24:units}{0.652\arcm}
\setsymbol{LFI:FWHM:uncertainty:LFI25:units}{1.075\arcm}
\setsymbol{LFI:FWHM:uncertainty:LFI26:units}{1.131\arcm}
\setsymbol{LFI:FWHM:uncertainty:LFI27:units}{1.266\arcm}
\setsymbol{LFI:FWHM:uncertainty:LFI28:units}{1.287\arcm}

\setsymbol{LFI:FWHM:uncertainty:LFI18}{0.170}
\setsymbol{LFI:FWHM:uncertainty:LFI19}{0.174}
\setsymbol{LFI:FWHM:uncertainty:LFI20}{0.170}
\setsymbol{LFI:FWHM:uncertainty:LFI21}{0.156}
\setsymbol{LFI:FWHM:uncertainty:LFI22}{0.164}
\setsymbol{LFI:FWHM:uncertainty:LFI23}{0.171}
\setsymbol{LFI:FWHM:uncertainty:LFI24}{0.652}
\setsymbol{LFI:FWHM:uncertainty:LFI25}{1.075}
\setsymbol{LFI:FWHM:uncertainty:LFI26}{1.131}
\setsymbol{LFI:FWHM:uncertainty:LFI27}{1.266}
\setsymbol{LFI:FWHM:uncertainty:LFI28}{1.287}


\setsymbol{HFI:center:frequency:100GHz:units}{100\,GHz}
\setsymbol{HFI:center:frequency:143GHz:units}{143\,GHz}
\setsymbol{HFI:center:frequency:217GHz:units}{217\,GHz}
\setsymbol{HFI:center:frequency:353GHz:units}{353\,GHz}
\setsymbol{HFI:center:frequency:545GHz:units}{545\,GHz}
\setsymbol{HFI:center:frequency:857GHz:units}{857\,GHz}

\setsymbol{HFI:center:frequency:100GHz}{100}
\setsymbol{HFI:center:frequency:143GHz}{143}
\setsymbol{HFI:center:frequency:217GHz}{217}
\setsymbol{HFI:center:frequency:353GHz}{353}
\setsymbol{HFI:center:frequency:545GHz}{545}
\setsymbol{HFI:center:frequency:857GHz}{857}


\setsymbol{HFI:Ndetectors:100GHz}{8}
\setsymbol{HFI:Ndetectors:143GHz}{11}
\setsymbol{HFI:Ndetectors:217GHz}{12}
\setsymbol{HFI:Ndetectors:353GHz}{12}
\setsymbol{HFI:Ndetectors:545GHz}{3}
\setsymbol{HFI:Ndetectors:857GHz}{4}


\setsymbol{HFI:FWHM:Maps:100GHz:units}{9\parcm88}
\setsymbol{HFI:FWHM:Maps:143GHz:units}{7\parcm18}
\setsymbol{HFI:FWHM:Maps:217GHz:units}{4\parcm87}
\setsymbol{HFI:FWHM:Maps:353GHz:units}{4\parcm65}
\setsymbol{HFI:FWHM:Maps:545GHz:units}{4\parcm72}
\setsymbol{HFI:FWHM:Maps:857GHz:units}{4\parcm39}
\setsymbol{HFI:FWHM:Maps:100GHz}{9.88}
\setsymbol{HFI:FWHM:Maps:143GHz}{7.18}
\setsymbol{HFI:FWHM:Maps:217GHz}{4.87}
\setsymbol{HFI:FWHM:Maps:353GHz}{4.65}
\setsymbol{HFI:FWHM:Maps:545GHz}{4.72}
\setsymbol{HFI:FWHM:Maps:857GHz}{4.39}


\setsymbol{HFI:beam:ellipticity:Maps:100GHz}{1.15}
\setsymbol{HFI:beam:ellipticity:Maps:143GHz}{1.01}
\setsymbol{HFI:beam:ellipticity:Maps:217GHz}{1.06}
\setsymbol{HFI:beam:ellipticity:Maps:353GHz}{1.05}
\setsymbol{HFI:beam:ellipticity:Maps:545GHz}{1.14}
\setsymbol{HFI:beam:ellipticity:Maps:857GHz}{1.19}


\setsymbol{HFI:FWHM:Mars:100GHz:units}{9\parcm37}
\setsymbol{HFI:FWHM:Mars:143GHz:units}{7\parcm04}
\setsymbol{HFI:FWHM:Mars:217GHz:units}{4\parcm68}
\setsymbol{HFI:FWHM:Mars:353GHz:units}{4\parcm43}
\setsymbol{HFI:FWHM:Mars:545GHz:units}{3\parcm80}
\setsymbol{HFI:FWHM:Mars:857GHz:units}{3\parcm67}

\setsymbol{HFI:FWHM:Mars:100GHz}{9.37}
\setsymbol{HFI:FWHM:Mars:143GHz}{7.04}
\setsymbol{HFI:FWHM:Mars:217GHz}{4.68}
\setsymbol{HFI:FWHM:Mars:353GHz}{4.43}
\setsymbol{HFI:FWHM:Mars:545GHz}{3.80}
\setsymbol{HFI:FWHM:Mars:857GHz}{3.67}


\setsymbol{HFI:beam:ellipticity:Mars:100GHz}{1.18}
\setsymbol{HFI:beam:ellipticity:Mars:143GHz}{1.03}
\setsymbol{HFI:beam:ellipticity:Mars:217GHz}{1.14}
\setsymbol{HFI:beam:ellipticity:Mars:353GHz}{1.09}
\setsymbol{HFI:beam:ellipticity:Mars:545GHz}{1.25}
\setsymbol{HFI:beam:ellipticity:Mars:857GHz}{1.03}


\setsymbol{HFI:CMB:relative:calibration:100GHz}{$\lsim 1\%$}
\setsymbol{HFI:CMB:relative:calibration:143GHz}{$\lsim 1\%$}
\setsymbol{HFI:CMB:relative:calibration:217GHz}{$\lsim 1\%$}
\setsymbol{HFI:CMB:relative:calibration:353GHz}{$\lsim 1\%$}
\setsymbol{HFI:CMB:relative:calibration:545GHz}{}
\setsymbol{HFI:CMB:relative:calibration:857GHz}{}


\setsymbol{HFI:CMB:absolute:calibration:100GHz}{$\lsim 2\%$}
\setsymbol{HFI:CMB:absolute:calibration:143GHz}{$\lsim 2\%$}
\setsymbol{HFI:CMB:absolute:calibration:217GHz}{$\lsim 2\%$}
\setsymbol{HFI:CMB:absolute:calibration:353GHz}{$\lsim 2\%$}
\setsymbol{HFI:CMB:absolute:calibration:545GHz}{}
\setsymbol{HFI:CMB:absolute:calibration:857GHz}{}


\setsymbol{HFI:FIRAS:gain:calibration:accuracy:statistical:100GHz}{}
\setsymbol{HFI:FIRAS:gain:calibration:accuracy:statistical:143GHz}{}
\setsymbol{HFI:FIRAS:gain:calibration:accuracy:statistical:217GHz}{}
\setsymbol{HFI:FIRAS:gain:calibration:accuracy:statistical:353GHz}{2.5\%}
\setsymbol{HFI:FIRAS:gain:calibration:accuracy:statistical:545GHz}{1\%}
\setsymbol{HFI:FIRAS:gain:calibration:accuracy:statistical:857GHz}{0.5\%}


\setsymbol{HFI:FIRAS:gain:calibration:accuracy:systematic:100GHz}{}
\setsymbol{HFI:FIRAS:gain:calibration:accuracy:systematic:143GHz}{}
\setsymbol{HFI:FIRAS:gain:calibration:accuracy:systematic:217GHz}{}
\setsymbol{HFI:FIRAS:gain:calibration:accuracy:systematic:353GHz}{}
\setsymbol{HFI:FIRAS:gain:calibration:accuracy:systematic:545GHz}{7\%}
\setsymbol{HFI:FIRAS:gain:calibration:accuracy:systematic:857GHz}{7\%}


\setsymbol{HFI:FIRAS:zero:point:accuracy:100GHz:units}{0.8\MJysr}
\setsymbol{HFI:FIRAS:zero:point:accuracy:143GHz:units}{}
\setsymbol{HFI:FIRAS:zero:point:accuracy:217GHz:units}{}
\setsymbol{HFI:FIRAS:zero:point:accuracy:353GHz:units}{1.4\MJysr}
\setsymbol{HFI:FIRAS:zero:point:accuracy:545GHz:units}{2.2\MJysr}
\setsymbol{HFI:FIRAS:zero:point:accuracy:857GHz:units}{1.7\MJysr}

\setsymbol{HFI:FIRAS:zero:point:accuracy:100GHz}{0.8}
\setsymbol{HFI:FIRAS:zero:point:accuracy:143GHz}{}
\setsymbol{HFI:FIRAS:zero:point:accuracy:217GHz}{}
\setsymbol{HFI:FIRAS:zero:point:accuracy:353GHz}{1.4}
\setsymbol{HFI:FIRAS:zero:point:accuracy:545GHz}{2.2}
\setsymbol{HFI:FIRAS:zero:point:accuracy:857GHz}{1.7}


\setsymbol{HFI:unit:conversion:100GHz:units}{0.2415\MJysrmK}
\setsymbol{HFI:unit:conversion:143GHz:units}{0.3694\MJysrmK}
\setsymbol{HFI:unit:conversion:217GHz:units}{0.4811\MJysrmK}
\setsymbol{HFI:unit:conversion:353GHz:units}{0.2883\MJysrmK}
\setsymbol{HFI:unit:conversion:545GHz:units}{0.05826\MJysrmK}
\setsymbol{HFI:unit:conversion:857GHz:units}{0.002238\MJysrmK}

\setsymbol{HFI:unit:conversion:100GHz}{0.2415}
\setsymbol{HFI:unit:conversion:143GHz}{0.3694}
\setsymbol{HFI:unit:conversion:217GHz}{0.4811}
\setsymbol{HFI:unit:conversion:353GHz}{0.2883}
\setsymbol{HFI:unit:conversion:545GHz}{0.05826}
\setsymbol{HFI:unit:conversion:857GHz}{0.002238}


\setsymbol{HFI:colour:correction:alpha=-2:V1.01:100GHz}{0.9893}
\setsymbol{HFI:colour:correction:alpha=-2:V1.01:143GHz}{0.9759}
\setsymbol{HFI:colour:correction:alpha=-2:V1.01:217GHz}{1.0007}
\setsymbol{HFI:colour:correction:alpha=-2:V1.01:353GHz}{1.0028}
\setsymbol{HFI:colour:correction:alpha=-2:V1.01:545GHz}{1.0019}
\setsymbol{HFI:colour:correction:alpha=-2:V1.01:857GHz}{0.9889}


\setsymbol{HFI:colour:correction:alpha=0:V1.01:100GHz}{1.0008}
\setsymbol{HFI:colour:correction:alpha=0:V1.01:143GHz}{1.0148}
\setsymbol{HFI:colour:correction:alpha=0:V1.01:217GHz}{0.9909}
\setsymbol{HFI:colour:correction:alpha=0:V1.01:353GHz}{0.9888}
\setsymbol{HFI:colour:correction:alpha=0:V1.01:545GHz}{0.9878}
\setsymbol{HFI:colour:correction:alpha=0:V1.01:857GHz}{1.0014}

\section{Introduction}

Very massive clusters above redshift $z\!\sim\!1$, when the Universe was at half the present age, are predicted to be very rare. 
They potentially provide a sensitive probe to constrain deviations from the standard $\Lambda$CDM paradigm \citep[e.g.][]{mor11}; e.g., owing to non-Gaussian perturbations, non-standard quintessence models or modified gravity models \citep[see][for a review]{all11}.
They are also ideal targets for studying key aspects of the gravitational physics that drives cluster formation, including measurement of the evolution of the mass concentration. For these reasons, the scientific community has, over the past two decades, put strong effort into the discovery and characterisation of these objects.

\begin{figure*}[t]
\begin{centering}
\begin{minipage}[t]{\textwidth}
\resizebox{\hsize}{!} {
\includegraphics[height=3cm,angle=0,keepaspectratio]{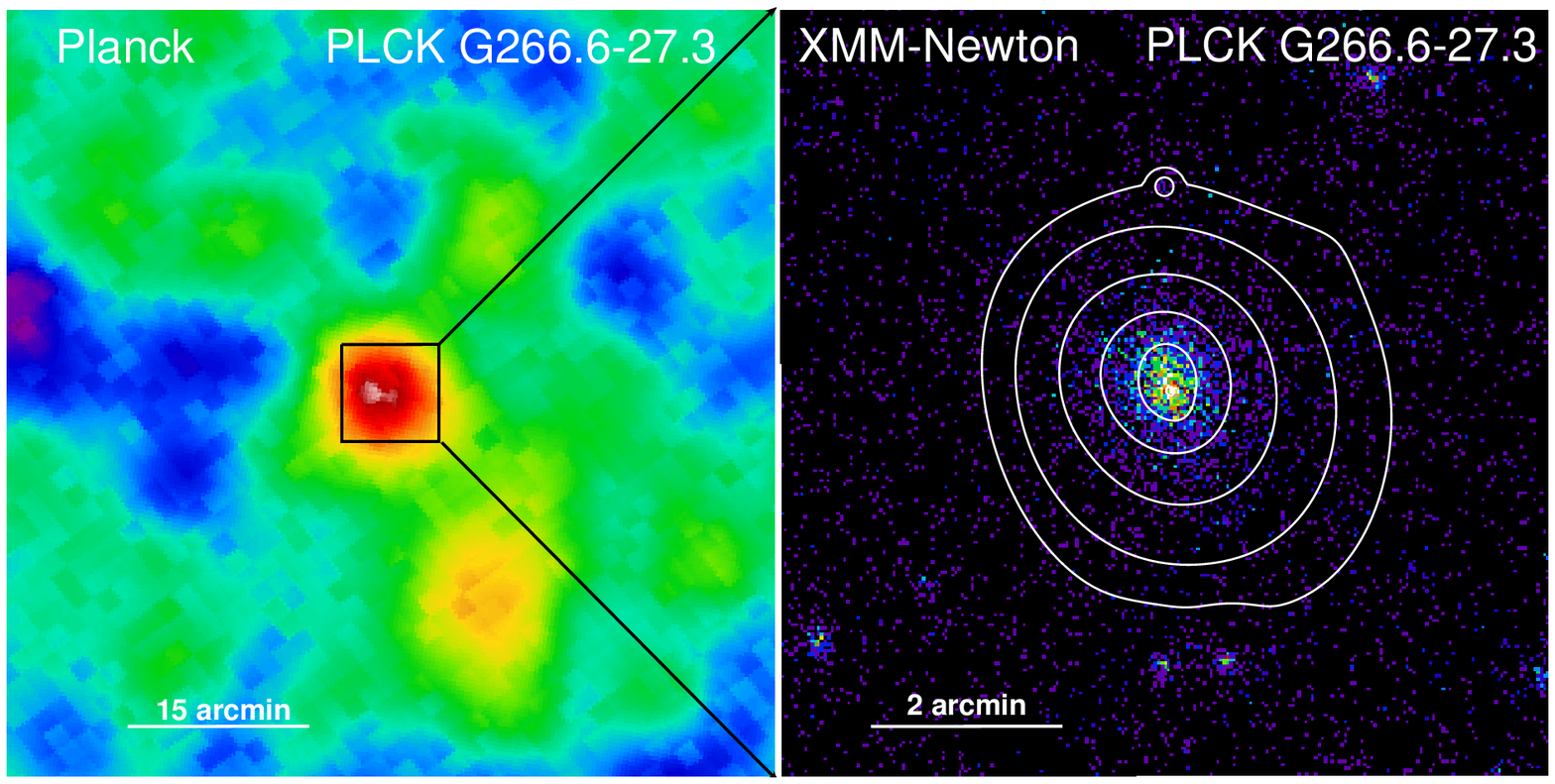}%
\hspace{2mm}
\includegraphics[height=3cm,angle=0,keepaspectratio]{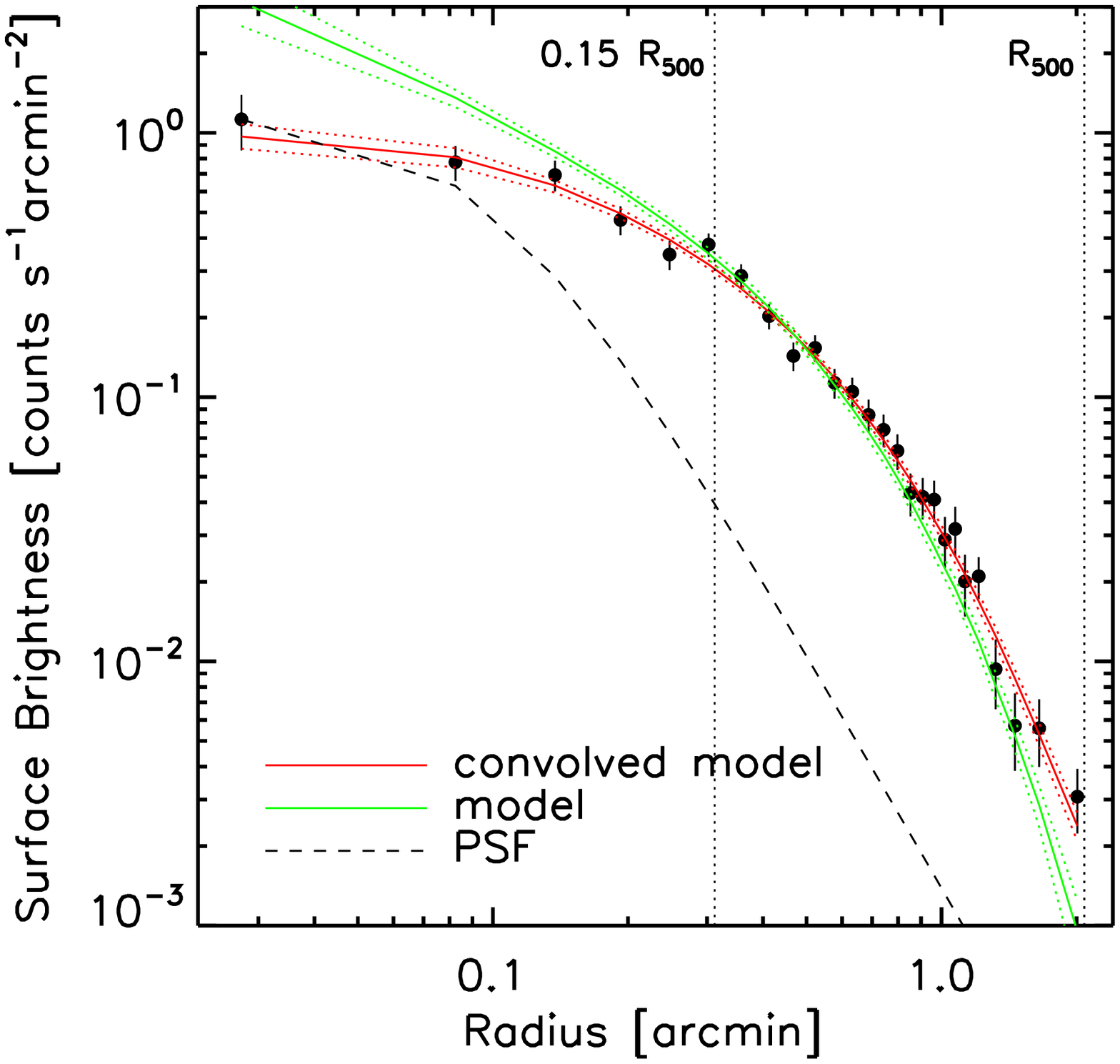}
}
\end{minipage}
\end{centering}
\caption{{\footnotesize Left panel:  \planck\ $Y_{\rm SZ}$ map of  \toto\ 
 obtained with the modified internal linear combination algorithm \citep[MILCA;][]{hur10} with a spatial resolution of $10\arcmin$. Middle panel:   \xmm\ exposure-corrected count rate image 
 of the region indicated by the black box in the left panel. It is 
 obtained using data from the \mos1$\&$2 and \pn\ camera in the $[0.3$--$2.0]\,\keV$ energy band. The contours of the \xmm\  image after wavelet filtering are overlaid in white. Right panel: corresponding \xmm\ surface brightness profile. The green line indicates the best-fitting $\beta$-model with a cusp (see text); the red line is this model convolved with the point spread function (PSF) of \xmm, and the dashed line is the on-axis PSF of \xmm, normalised to the central intensity. The source is clearly significantly extended. }}\label{fig:ima}
\end{figure*}

Until recently it was possible to identify clusters of galaxies only via optical/infrared or X-ray surveys. Indeed, the most distant clusters presently known have all been detected with these techniques, e.g., the IR-selected cluster CL\,J1449+0856 at $z\!\!=\!\!2.07$ \citep{gob11} and the X-ray selected system XMMU\,J105324.7+572348 at $z\!\!=\!\!1.75$ \citep{hen10}. 
For both of these objects, extended X-ray emission has been detected with \xmm, confirming their status as fully established galaxy clusters; however, their total masses are more typical of systems in the poor cluster or group regime ($\lesssim 10^{14}\, \msol$).
Until recently, the most massive cluster known in the $z\gtrsim1$ universe was  XMMU\,J2235.3$-$2557 at $z\!\!=\!\!1.39$, discovered in the \xmm\ Distant Cluster Project (XDCP) based on serendipitous cluster searches in \xmm\ observations \citep{mul05}. For this system, \citet{jee09} estimate a mass of $M_{200}=(7.3\pm1.3)\times10^{14}\,\msol$ from a weak lensing analysis. 

However, clusters are also detectable through the Sunyaev-Zeldovich (SZ) effect \citep{sun72}, the spectral distortion of the cosmic microwave background (CMB) generated via inverse Compton scattering of CMB photons by the hot electrons in the intra-cluster medium. Crucially, the total SZ signal is expected to be closely related to the cluster mass \citep[e.g.][]{das04}, and its brightness insensitive to redshift dimming. As a result, SZ surveys can potentially provide unbiased cluster samples that are as close as possible to being mass-selected\footnote{In practice, the mass threshold detectable by \planck\ increases with redshift. The total SZ signal is not resolved by \planck\ at high $z$ and it decreases with $z$ due to the decreasing angular size of the object.}. They offer an ideal way to identify massive, high-redshift clusters. One recent illustration is the detection of SPT-CL\,J2106$-$5844  at $z=1.13$ by the South Pole Telescope (SPT) survey \citep{fol11}. With an estimated mass of  $M_{200}=(1.27\pm0.21)\times10^{15}\,\msol$, SPT-CL\,J2106$-$5844  is nearly twice as  massive as XMMU\,J2235.3$-$2557. 
 
 The \planck\footnote{\Planck\ (http://www.esa.int/Planck) is a project of the European Space Agency (ESA) with instruments provided by two scientific consortia funded by ESA member states (in particular the lead countries: France and Italy) with contributions from NASA (USA), and telescope reflectors provided in a collaboration between ESA and a scientific consortium led and funded by Denmark.} satellite has been surveying the sky in the microwave band since August 2009 \citep{planck2011-1.1} with a good (band-dependent) spatial resolution of 5 arcmin \citep{planck2011-1.4,planck2011-1.5}.  Compared to other SZ experiments such as ACT \citep{mar11} or SPT \citep{car09a}, \planck\ brings unique nine-band coverage from 30 to 857 GHz and, most crucially, an exceptionally large survey volume.  \planck\ is the first all-sky survey capable of blindly detecting clusters (i.e.,  not guided in the search by prior observations), since the \rosat\  All-Sky Survey  (RASS) in the X-ray domain.   This coverage allows detection of the rarest clusters, the most massive objects lying in the exponential tail of the mass functitheron.
 
\citet{planck2011-5.1a} recently published the Early SZ (ESZ) sample, the first sample of galaxy clusters detected blindly in the all-sky maps from the first ten months of the \planck\ survey. The properties of this first sample already show that \planck\ is detecting previously unknown, massive clusters that do not appear in \rass\ or in other smaller area SZ surveys \citep{planck2011-5.1b}. The ESZ comprises high signal-to-noise ratio  ($\textrm{S/N}>6$)  \planck\ SZ sources up to $z=0.5$.  We report here on an SZ source that was blindly identified at $\textrm{S/N}\sim5$ in the \planck\ all-sky survey, and recent \xmm\ validation observations confirm it is a massive cluster at $z\sim1$.

In this paper, we adopt a $\Lambda$CDM cosmology with $H_0=70\,\kmsMpc$, $\Omega_{\rm M}=0.3$, and $\Omega_\Lambda=0.7$. The factor $E(z)= \sqrt{\Omega_{\rm M} (1+z)^3+\Omega_\Lambda}$ is the ratio of the Hubble constant at redshift $z$ to its present-day value. The quantities $M_{\rm \delta}$ and $R_{\rm \delta}$ are the total mass and radius corresponding to a total density contrast $\delta$, as compared to $\rho_{\rm c}(z)$, the critical density of the Universe at the cluster redshift;  $M_{\rm \delta} = (4\pi/3)\,\delta\,\rho_c(z)\,R_{\rm \delta}^3$. The SZ flux is characterised by $\YSZ$, where $\YSZ\,D_{\rm A}^2$ is the spherically integrated Compton parameter within $\Rv$ (corresponding to $\delta=500$), and $D_{\rm A}$ is the angular-diameter distance to the cluster. 

\section{\textit{Planck} detection}

The  blind search for clusters in \planck\ data relies on a multi-matched filter (MMF) approach \citep{mel06}.  Candidates then undergo a validation process, including internal quality checks  and cross-correlation with ancillary data and catalogues, as described in \citet{planck2011-5.1a}.  This process produces a list of new \planck\ SZ cluster candidates above a given $\textrm{S/N}$ threshold that require follow-up for confirmation.  The  \xmm\  follow-up for validation, undertaken in Director's Discretionary Time via an agreement between the \xmm\ and \planck\ Project Scientists, plays a central role in this confirmation procedure. It consists of snapshot  exposures  ($\sim10\,{\rm ks}$), sufficient for unambiguous discrimination between clusters and false candidates \citep{planck2011-5.1b}. The results of the first two runs (completed in September 2010)  are reported by \citet{planck2011-5.1a,planck2011-5.1b}.  

The \xmm\ validation programme is continuing to explore lower $\textrm{S/N}$ and detection quality criteria. \toto,  detected at  $\textrm{S/N}=5.03$, was observed in the framework of the third run of the \xmm\ validation programme, for which the analysis is on-going. This run comprises a  total of 11 candidates detected at  $4.5 < \textrm{S/N} < 5.3$  from the same \planck\ HFI maps used for the construction the ESZ sample.  The 11 candidates were sent for scheduling in November 2010 and the observations were performed between  22 December 2010 and 16 May 2011. Interestingly, \toto\ has been independently detected in the SPT survey.  Its  \planck\ position 
($6^{\rm h}\,16^{\rm m}\,6\fs6$, $-57^{\circ}\,47\arcmin\,29\arcs$) 
 is consistent with that  of SPT-CL\,J0615-5746 \citep[][published on \url{arXiv.org} in January 2011, with  a photometric redshift of  $z_{\rm phot}=1\pm0.1$]{wil11}.

\begin{figure}[t]
\center
\includegraphics[scale=1.,angle=-90,keepaspectratio,width=0.8\columnwidth]{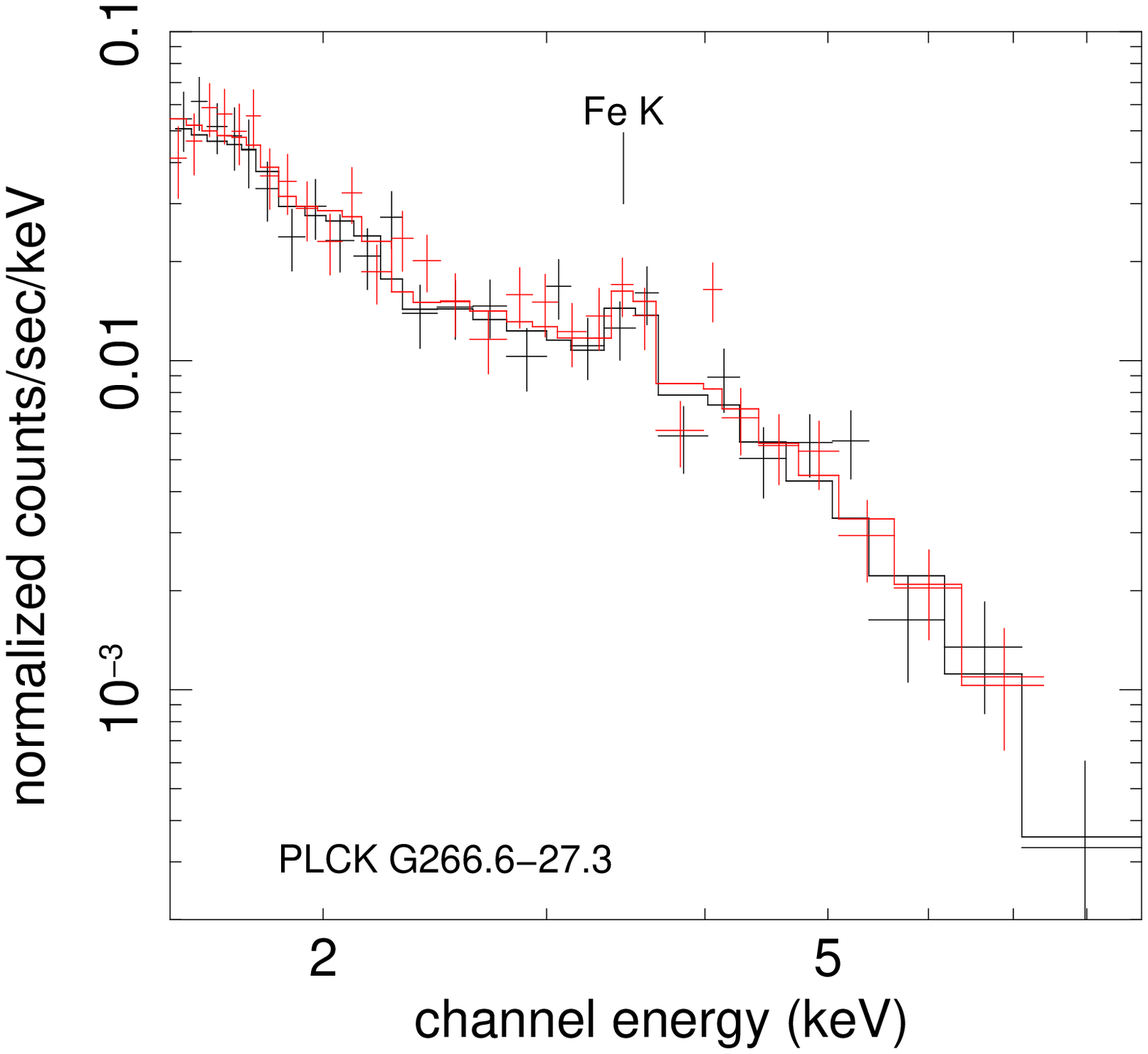}
\caption{{\footnotesize \xmm\  \mos1 (black) and \mos2 (red)  spectra extracted from a circular region of $1\farcm5$ in radius and centred in the cluster X-ray peak. Only the data points above $2\,\keV$ are shown for clarity, but data down to $0.3\,\keV$ are used in the spectral fitting. The line is the thermal model for the best-fitting redshift, $z = 0.94 \pm  0.02$. The position of the redshifted Fe K line is marked.}}\label{fig:spec}
\end{figure}

\section{\textit{XMM-Newton} validation}
	
\subsection{Observation and data reduction}

\toto\  was observed with the \xmm\ EPIC instrument \citep{tur01,str01}, using the thin filters and the extended full frame mode for the `pn-CCD' camera.  The data analysis and validation procedure is described in \citet{planck2011-5.1b}.  Calibrated event lists were produced with v11.0 of the \xmm\ Science Analysis System.  Data that are affected by periods of high background due to soft proton flares were omitted from the analysis, and  the remaining data were  {\sc pattern}-selected and corrected for vignetting, as described in \citet{pra07}. Bright point sources were excised from the data. 
Background treatment is described in \citet{pra10}. In the spectroscopic analysis, the cluster component was modelled with an absorbed thermal emission model (\mekal) with a hydrogen column density fixed at the 21-cm value of \citet{dic90}. 

The observation, $\textrm{OBSID}=0658200101$, was affected by soft proton flares. The net exposure time after flare cleaning is  only $2.4\,{\rm ks}$ for the pn-CCD camera, with a particle background  $30\%$ higher than nominal. The MOS camera data are less affected with a clean time of $\sim12\,{\rm ks}$  and a background excess about two times lower. We undertook a conservative approach to analysing  spectroscopic data, since they are  the most sensitive to the background estimate. We first fitted the data from the three cameras simultaneously, then fitted only the MOS cameras. The uncertainties in the physical quantities below reflect the difference in best-fitting values between the two analyses and their errors.
 
\subsection{Confirmation and $z$ estimate}

In Fig.~\ref{fig:ima} we show the vignetting-corrected count rate image of the cluster in the $[0.3$--$2.0]\,\keV$ band.  An extended X-ray source is clearly coincident with \toto. Its total EPIC count rate in the $[0.3$--$2.0]\,\keV$ band  is  $(0.52\pm0.02)\,{\rm count/s}$ within $2.3\arcmin$,  the maximum radius of detection. The offset between the X-ray cluster centre, defined as the emission peak at $6^{\rm h}\,15^{\rm m}\,51\fs7$, $-57^{\circ}\,46\arcmin 52\farcs8$, and  the \planck\ blind position is  $2\farcm07$, consistent with the position reconstruction uncertainty, driven by the \planck\ spatial resolution and the source ${\rm S/N}$ \citep{planck2011-5.1a}. The extended nature of the source is confirmed  by comparing  the  surface brightness profile with the \xmm\ point spread function (PSF) (Fig.~\ref{fig:ima}, right panel).  A typical (PSF-convolved) cluster surface brightness model consisting of a $\beta$-model with a central cusp \citep[Eq. 2 in ][]{pra02} provides a good fit to the data  and further supports the extended nature of the source (Fig.~\ref{fig:ima}). 

We extracted a spectrum within a circular region corresponding to the maximum significance of the X--ray detection ($\theta \lesssim 1\farcm5$). The iron K line complex is clearly detected (Fig.~\ref{fig:spec}).  Its  significance is $3.6\sigma$, estimated from a fit of the spectrum in the $[2$--$6]\,\keV$ band with a continuum plus a Gaussian line model.  Since the  centroid of the line complex  depends on the temperature, the redshift is determined from a thermal model fit to the full spectrum,  as described in detail in \citet{planck2011-5.1b}. This  yields a precise redshift estimate of $z=0.94\pm 0.02$.

\section{Physical cluster properties}

\subsection{An exceptionally luminous and massive  cluster}

We derived the deprojected, PSF-corrected gas density profile  from the surface brightness profile, using the non-parametric method described  in \citet{cro06}. Global cluster parameters were then estimated self-consistently within $\Rv$ via iteration about the $\Mv$--$\YX$ relation of \citet{arn10}, assuming standard evolution,  $E(z)^{2/5}\Mv = 10^{14.567 \pm 0.010} \left[\frac{\YX}{2\times10^{14}\,{\msol}\,\keV}\right]^{0.561 \pm 0.018}\,\msol$. The quantity $\YX$, introduced by \citet{kra06},  is defined as the product  of  $\Mgv$, the gas mass within $\Rv$, and $\TX$.  $\TX$ is the spectroscopic temperature measured in the $[0.15$--$0.75]\,\Rv$ aperture.
 In addition, $L_{500}$, the X-ray luminosity inside $\Rv$, was calculated as described in \citet{pra09}. All resulting  X-ray properties, including iron abundance, are summarised in Table~\ref{tab:par}. 
 
\toto\ is an exceptionally luminous system. Its $[0.1$--$2.4]\,\keV$ band luminosity of $(22.7 \pm 0.8) \times10^{44}\,{\rm erg\,s^{-1}}$ is equal to that of  the fifth most luminous object in the MCXC compilation
of \citet{pif10},  MACS\,J0717.5+3745 at $z=0.55$, discovered in the \rass\ by \citet{edg03}. Moreover,  its $[0.5$--$2.0]\,\keV$ band luminosity is consistent with that of SPT-CL\,J2106$-$5844, the most luminous cluster known beyond $z=1$ \citep{fol11}.
Collectively, these three clusters are the most luminous systems at $z > 0.5$. They are only $40\%$  fainter than RXJ\,1347.5-1145, the most X-ray luminous cluster known in the Universe \citep{pif10}.


\begin{table} [t]                 
\begingroup
\newdimen\tblskip \tblskip=5pt
\caption{Physical properties of PLCK\,G266.6$-$27.3 derived from \textit{XMM-Newton} data.   \label{tab:par} } 
\label{FILL THIS IN}
\nointerlineskip
\vskip -3mm
\footnotesize
\setbox\tablebox=\vbox{
   \newdimen\digitwidth
   \setbox0=\hbox{\rm 0}
   \digitwidth=\wd0
   \catcode`*=\active
   \def*{\kern\digitwidth}
   \newdimen\signwidth
   \setbox0=\hbox{+}
   \signwidth=\wd0
   \catcode`!=\active
   \def!{\kern\signwidth}
\openup 2pt
\halign{\hbox to 1.4in{#\leaderfil}\tabskip=2em&
        \hfil#\hfil\tabskip=0pt\cr
\noalign{\doubleline}
\omit\hfil Parameter\hfil&Value\cr
\noalign{\vskip 3pt\hrule\vskip 5pt}
$z$&                       $0.94\pm0.02$\cr
Abundance&                 $0.44\pm0.17$\,solar\cr
$R_{500}$&                 $0.98\pm0.03$\,Mpc\cr
$M_{500}$&                 $7.8^{+0.8}_{-0.7} \times 10^{14}$\,\sol\cr
$Y_{\rm X}$&               $1.10^{+0.20}_{-0.17} \times 10^{15}$\,\sol\,keV\cr
$T_{\rm X}$&               $10.5^{+1.6}_{-1.4}$\,keV\cr
$T(<R_{500})$&             $11.4^{+1.4}_{-1.2}$\,keV\cr
$L_{500}$([0.5--2.0]\,keV)&$14.2\pm0.5 \times 10^{44}$\,erg\,s\mo\cr
$L_{500}$([0.1--2.4]\,keV)&$22.7\pm0.8 \times 10^{44}$\,erg\,s\mo\cr
\noalign{\vskip 5pt\hrule\vskip 3pt}}}
\endPlancktable
\endgroup
\tablefoot{
The $M_{500}$-$Y_{\rm X}$ relation with self-similar evolution is used to estimate $R_{500}$ (see Sect.~4). 
}
\end{table}

Consistent with expectations for high-redshift \planck-detected clusters, we find that  this cluster is extremely hot, $\TX\sim11\,\keV$, and massive, with a mass of  $\Mv = 7.8_{-0.7}^{+0.8}\times10^{14}\,\msol$. Our mass estimate is consistent with the less precise value, $\Mv=8\pm2\,{\rm (statistical)}\pm1.9\,{\rm (systematic)}\times10^{14}\,\msol$, which is derived by \citet{wil11} using the relation between SPT S/N and mass.
Comparison of  the masses of high-redshift systems is not trivial, because the estimation  strongly depends on method, e.g which mass proxy is used and at what  reference radius the mass is measured.  On the basis of the published mass estimates, \toto\  would appear to be the most massive cluster at $z\sim 1$. Using the same factor to convert $\Mv$ to $M_{200}$ as \citet{fol11}, we obtain $M_{200}=15.5_{-1.4}^{+1.5} \times10^{14}\,\msol$, to be compared to $M_{200}=(12.7\pm2.1)\footnote{The error includes an extra $\sim 15\%$ error accounting for uncertainties in the scaling relations.}  \times10^{14}\,\msol$ for SPT-CL\,J2106-5844. However, the last value was derived by combining X-ray and SZ  data. A more direct comparison of  $\Mv$ values estimated from the \MYX\ relation indicates that they are identical within their uncertainties: $\Mv=(7.8\pm0.8)\times10^{14}\,\msol$ for \toto\  and $\Mv=(9.3\pm2.0) \times 10^{14}\,\msol$ for SPT-CL\,J2106$-$5844.

\subsection{$\YSZ$ Compton parameter versus $\YX$}

The MMF blind detection was performed using the universal pressure profile of \citet{arn10} as a spatial template, leaving the position, characteristic size, $\theta_{500}$, and SZ flux, $\YSZ$,  as free parameters. The resulting flux, $\YSZ=(5.6  \pm 3.0)\times 10^{-4}\,{\rm arcmin}^2$, is consistent with the value,  $\YSZ^{\rm X}=6.4_{-1.0}^{+1.2}\times10^{-4} {\rm arcmin}^2$,  expected from the measured value of $\YX$ using the scaling relation derived from the universal pressure profile \citep[][Eq.~19]{arn10}.  The cluster size, comparable to \planck's spatial resolution, is poorly constrained, $\theta_{500}=3\farcm3\pm2\farcm8$. As discussed in \citet{planck2011-5.1a}, the uncertainty on the blind $\YSZ$ value is then large because of the flux-size degeneracy, where an overestimate of the  cluster size induces an overestimate of the SZ signal. The SZ photometry can be improved by using the more precise \xmm\ position and size in the flux extraction. The $\YSZ$ value  obtained using these X-ray priors,  $\YSZ=(4.1 \pm 0.9)\times 10^{-4}\,{\rm arcmin}^2$,  is lower than the value expected from the X-ray data at the  $1.7\sigma$ significance level. 

\begin{table}[t]
\centering  
\caption{ {\footnotesize SZ flux derived from \planck\ data with the reference value indicated in boldface.}
 \label{tab:parsz}}
\begin{tabular}{lccc}
\toprule
\toprule
Method  &  Definition  &  Value  &  $\theta_{500}$   \\
&                                     & ($10^{-4}\,{\rm arcmin}^2$) & (arcmin)   \\
\midrule
MMF blind & $\YSZ$ & $5.6  \pm 3.0$&$3.3\pm2.8$\\
PWS blind & $\YSZ$ & $6.5  \pm 1.8$&$3.9\pm1.6$\\
{\bf MMF X-ray prior}& $\YSZ$ & ${\bf 4.1  \pm 0.9 }$ & fixed\\
PWS X-ray prior & $\YSZ$ &  $5.3  \pm 0.9 $ & fixed \\
 MILCA &  $Y_{\rm tot}$ & $5.9  \pm 1.0$ & \dots \\
\bottomrule
\end{tabular}
\tablefoot{
Uncertainties on the blind values take into account the size uncertainty.}
\end{table}

To check the robustness of the $\YSZ$ estimate, we compared  the MMF value with the one derived from the PowellSnakes \citep[PWS;][]{car09b,car11} algorithm and  the modified internal linear combination algorithm \citep[MILCA;][]{hur10}. The  values are given in Table~\ref{tab:parsz}.  PWS is a  blind detection algorithm that assumes the same profile shape as MMF,  but is based on a Bayesian statistical approach, as fully described in \citet{car09b}.  MMF and PWS give consistent results, the difference between MMF and PWS $\YSZ$ values being about 1.3 times the respective $1\sigma$ uncertainties.  MILCA  is a component separation method that allows reconstruction of the SZ map around the cluster from an optimised linear combination of \planck\  HFI maps. In contrast to the MMF and PWS methods, the SZ flux derived from MILCA is obtained from aperture photometry, i.e., with no assumptions on SZ profile shape or size. Assuming a typical conversion factor of $2/3$ based on the universal profile to convert the total $Y_{\rm tot}$ MILCA measurement  to $\YSZ$, the MMF and MILCA estimates are in excellent agreement.

Several factors may affect the X-ray and SZ flux measurements and bring them out of accord. We have checked for possible AGN contamination that could lower the $\YSZ$ value using the NVSS (at 1.4\,GHz, \citealt{con98}) and SUMSS (at 0.84\,GHz, \citealt{boc99}) catalogues, but no bright radio sources are found in the cluster vicinity. The closest radio source with significant flux density is at $12\farcm6$ away. The source has a 1\,GHz flux density of 0.46\,Jy. We also find no evidence of radio contamination in the low-frequency \planck\ bands. 
On the other hand, the $\YX$ measurement may also be increased by AGN  contamination, from  cluster members or foreground/background galaxies. Point source contamination is difficult to estimate owing to the \xmm\ PSF. So, we estimate a maximum contribution to the X-ray luminosity  from a central active galaxy of $\lesssim 20\%$, assuming a point source model normalised to the central value of the X-ray surface brightness. The contribution to the gas mass, hence  to $\YX$, would be less, provided that the source is not hard enough to significantly affect $\TX$.  Nevertheless, only high-resolution X-ray imaging (e.g., from Chandra) can definitively establish whether X-ray AGNs at the cluster location affect our luminosity or mass measurement.  A departure from the universal pressure profile would change the $\YSZ/\YX$ ratio. The density profile shown below does not show any indication of this effect;  however, deep spatially resolved \xmm\ and \chandra\  spectroscopic observations are needed to derive the radial pressure gradient from the core to $\Rv$. A final interesting possibility is that gas clumping could affect the $\YX$ measurements.  A combination of X-ray and higher resolution SZ observations is required to assess this point.

\begin{figure}[t]
\center
\includegraphics[scale=1.,angle=0,keepaspectratio,width=0.9\columnwidth]{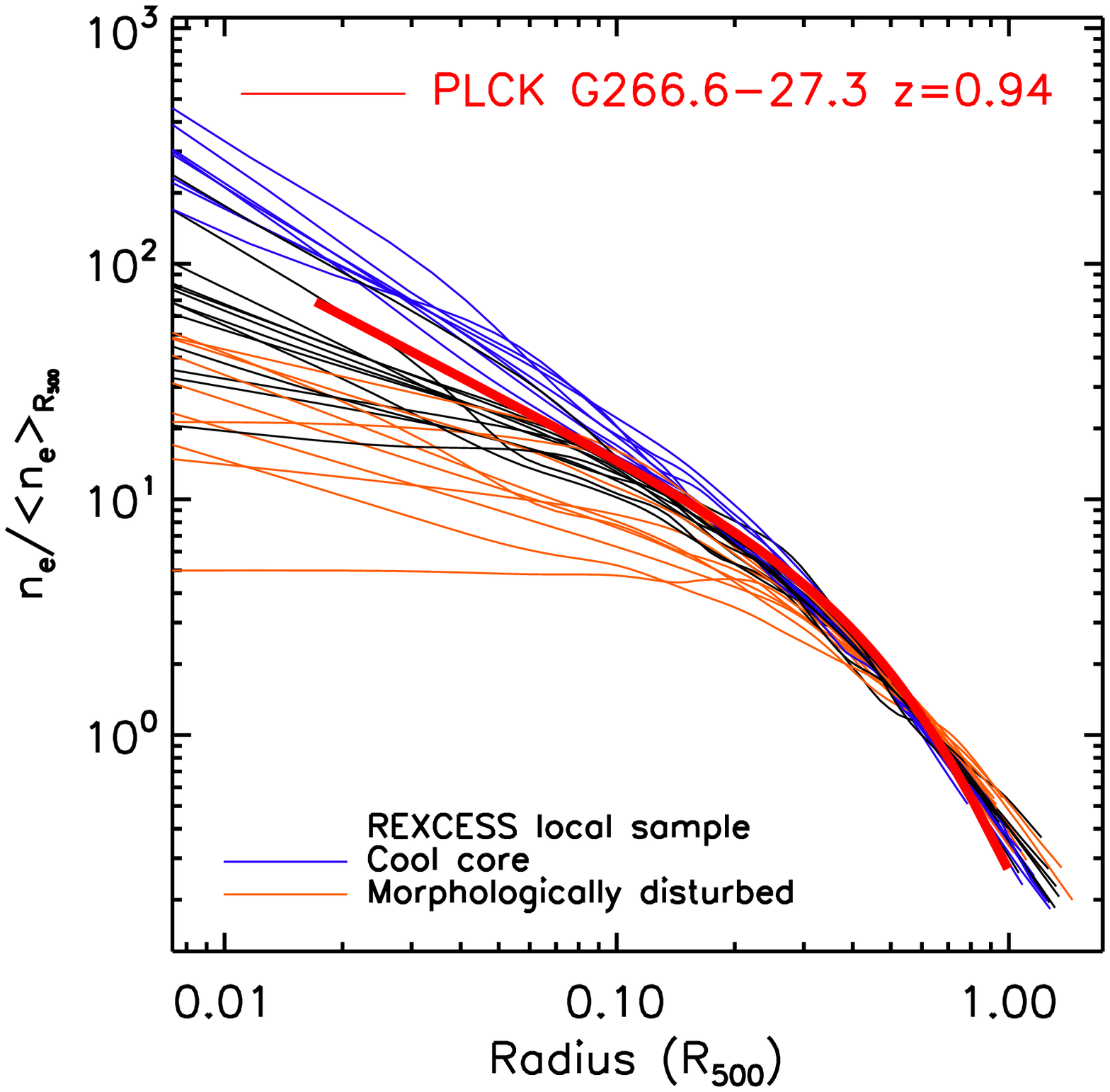}
\caption{{\footnotesize Scaled density profiles of the \rexcess\ local cluster sample \citep{boe07,cro08}. The blue lines show the profiles for the cool core systems, and the orange lines  the density profiles of the morphologically disturbed systems. The density profile of \toto\ is shown with a thick red  line.}}\label{fig:profiles}
\end{figure}

\subsection{Dynamical state and self-similarity of shape up to high z}
The available information indicates that \toto\ may be particularly dynamically relaxed. The cluster image  (Fig.~\ref{fig:ima}, middle panel) does not show any sign of disturbance: the surface brightness is quite regular and quasi-azimuthally symmetric within $R_{500}$. The offset between the X-ray surface brightness peak and  the  cluster brightest galaxy \citep[][Fig.19]{wil11} is less than $5\arcsec$. 

To further examine the dynamical state of the cluster, in Fig.~\ref{fig:profiles}  we compare its scaled density profile to those of clusters in  the local representative X-ray-selected sample \rexcess\ \citep{boe07,cro08}. The radii are scaled by $R_{500}$ and the density by the mean within $\Rv$.
As extensively discussed by \citet{pra09} and \citet{arn10},  morphologically disturbed (i.e., merging) systems have systematically
shallower density profiles than more relaxed cool core objects. This is illustrated in Fig.~\ref{fig:profiles}, where we indicate the scaled density profiles of the more relaxed cool core  and
the dynamically active merging clusters. 
The scaled density profile of \toto\ lies between the two classes,  but with an indication of being closer to the relaxed rather than the merging systems. It is thus possible that \toto\  is a cool core object at $z\sim 1$. 
Such objects are expected to be rare \citep[e.g.,][]{vik07,san10}, and no cluster at this redshift has yet been found to contain a resolved central temperature drop that would confirm the presence of a cool core. A deep exposure at \chandra\ spatial resolution  is needed to check this hypothesis.

It is worth emphasising the similarity beyond the core of the density profile of this cluster with respect to \rexcess\ systems. This is the first piece of evidence for a similarity of shape up to redshifts as high as $z\sim1$. 

\section{Conclusion}
	
\toto\  is the first  blindly discovered \planck\  cluster of galaxies at $z\sim1$. It has been confirmed by \xmm\ in the framework of the on-going validation DDT observations. \xmm\ data allowed us to measure the redshift with high accuracy ($z = 0.94\pm0.02$) and estimate the cluster mass to be $\Mv = (7.8\pm0.8) \times 10^{14}\,\msol$. This \xmm\ confirmation and redshift estimate is a clear  demonstration of the capability of Planck for detecting high-$z$, high-mass clusters.

\toto\ is an exceptional system, both in terms of its luminosity and its estimated mass. Furthermore, unlike other high-redshift clusters, it  is likely to be a relaxed system,  potentially allowing accurate hydrostatic mass measurements. It is thus a perfect target for deep multi-wavelength follow-up to address such important cosmological issues as the evolution of dark matter profiles, the evolution of the mass-$Y_{\rm SZ}$ relation, gas clumping,  and the bias between X-ray and lensing mass estimates at such high redshift.

\begin{acknowledgements}
  The \planck\ Collaboration thanks Norbert Schartel for his
  support of the validation process and for granting discretionary time for the observation of \planck\ cluster candidates.  
  The present work is based 
  on observations obtained with \xmm, an ESA science mission with
  instruments and contributions directly funded by ESA Member States
  and the USA (NASA). This research has made use of the following databases: SIMBAD, operated at the CDS, Strasbourg, France; the NED database, which is operated by the Jet Propulsion Laboratory, California Institute of Technology, under contract with the National Aeronautics and Space Administration; BAX, which is operated by the Laboratoire dÕAstrophysique de Tarbes-Toulouse (LATT), under contract with the Centre National dÕEtudes Spatiales (CNES); and the  SZ repository operated by
IAS Data and Operation Center (IDOC) under contract with CNES.
A description of the Planck Collaboration and a
list of its members, indicating which technical or scientific
activities they have been involved in, can be found at  http://www.rssd.esa.int/Planck\_Collaboration.
The Planck Collaboration acknowledges the support of: ESA; CNES and CNRS/INSU-IN2P3-INP (France); ASI, CNR, and INAF (Italy); NASA and DoE (USA); STFC and UKSA (UK); CSIC, MICINN and JA (Spain); Tekes, AoF and CSC (Finland); DLR and MPG (Germany); CSA (Canada); DTU Space (Denmark); SER/SSO (Switzerland); RCN (Norway); SFI (Ireland); FCT/MCTES (Portugal); and DEISA (EU)

\end{acknowledgements}

\bibliographystyle{aa}
\bibliography{Planck_plckg266.bib,Planck_bib.bib}

\begin{thebibliography}{39}
\expandafter\ifx\csname natexlab\endcsname\relax\def\natexlab#1{#1}\fi

\bibitem[{{Allen} {et~al.}(2011){Allen}, {Evrard}, \& {Mantz}}]{all11}
{Allen}, S.~W., {Evrard}, A.~E., \& {Mantz}, A.~B. 2011, \araa\ in press,
  arXiv:astro-ph/1103.4829

\bibitem[{{Arnaud} {et~al.}(2010){Arnaud}, {Pratt}, {Piffaretti},
  {B{\"o}hringer}, {Croston}, \& {Pointecouteau}}]{arn10}
{Arnaud}, M., {Pratt}, G.~W., {Piffaretti}, R., {et~al.} 2010, \aap, 517, A92

\bibitem[{{Bock} {et~al.}(1999){Bock}, {Large}, \& {Sadler}}]{boc99}
{Bock}, D., {Large}, M.~I., \& {Sadler}, E.~M. 1999, \aj, 117, 1578

\bibitem[{{B{\"o}hringer} {et~al.}(2007){B{\"o}hringer}, {Schuecker}, {Pratt},
  {Arnaud}, {Ponman}, {Croston}, {Borgani}, {Bower}, {Briel}, {Collins},
  {Donahue}, {Forman}, {Finoguenov}, {Geller}, {Guzzo}, {Henry}, {Kneissl},
  {Mohr}, {Matsushita}, {Mullis}, {Ohashi}, {Pedersen}, {Pierini}, {Quintana},
  {Raychaudhury}, {Reiprich}, {Romer}, {Rosati}, {Sabirli}, {Temple}, {Viana},
  {Vikhlinin}, {Voit}, \& {Zhang}}]{boe07}
{B{\"o}hringer}, H., {Schuecker}, P., {Pratt}, G.~W., {et~al.} 2007, \aap, 469,
  363

\bibitem[{{Carlstrom} {et~al.}(2009){Carlstrom}, {Ade}, {Aird}, {Benson},
  {Bleem}, {Busetti}, {Chang}, {Chauvin}, {Cho}, {Crawford}, {Crites}, {Dobbs},
  {Halverson}, {Heimsath}, {Holzapfel}, {Hrubes}, {Joy}, {Keisler}, {Lanting},
  {Lee}, {Leitch}, {Leong}, {Lu}, {Lueker}, {McMahon}, {Mehl}, {Meyer}, {Mohr},
  {Montroy}, {Padin}, {Plagge}, {Pryke}, {Ruhl}, {Schaffer}, {Schwan},
  {Shirokoff}, {Spieler}, {Staniszewski}, {Stark}, \& {Vieira}}]{car09a}
{Carlstrom}, J.~E., {Ade}, P.~A.~R., {Aird}, K.~A., {et~al.} 2009, {\tt
  arXiv:0907.4445}

\bibitem[{{Carvalho} {et~al.}(2009){Carvalho}, {Rocha}, \& {Hobson}}]{car09b}
{Carvalho}, P., {Rocha}, G., \& {Hobson}, M.~P. 2009, \mnras, 393, 681

\bibitem[{{Carvalho} {et~al.}(2011){Carvalho}, {Rocha}, {Hobson}, \&
  {Lasenby}}]{car11}
{Carvalho}, P., {Rocha}, G., {Hobson}, M.~P., \& {Lasenby}, A. 2011, to be
  submitted to \mnras

\bibitem[{{Condon} {et~al.}(1998){Condon}, {Cotton}, {Greisen}, {Yin},
  {Perley}, {Taylor}, \& {Broderick}}]{con98}
{Condon}, J.~J., {Cotton}, W.~D., {Greisen}, E.~W., {et~al.} 1998, \aj, 115,
  1693

\bibitem[{{Croston} {et~al.}(2006){Croston}, {Arnaud}, {Pointecouteau}, \&
  {Pratt}}]{cro06}
{Croston}, J.~H., {Arnaud}, M., {Pointecouteau}, E., \& {Pratt}, G.~W. 2006,
  \aap, 459, 1007

\bibitem[{Croston {et~al.}(2008)Croston, Pratt, B{\"o}hringer, Arnaud,
  Pointecouteau, Ponman, Sanderson, Temple, Bower, \& Donahue}]{cro08}
Croston, J.~H., Pratt, G.~W., B{\"o}hringer, H., {et~al.} 2008, \aap, 487, 431

\bibitem[{da~Silva {et~al.}(2004)da~Silva, Kay, Liddle, \& Thomas}]{das04}
da~Silva, A.~C., Kay, S.~T., Liddle, A.~R., \& Thomas, P.~A. 2004, \mnras, 348,
  1401

\bibitem[{{Dickey} \& {Lockman}(1990)}]{dic90}
{Dickey}, J.~M. \& {Lockman}, F.~J. 1990, \araa, 28, 215

\bibitem[{{Edge} {et~al.}(2003){Edge}, {Ebeling}, {Bremer}, {R{\"o}ttgering},
  {van Haarlem}, {Rengelink}, \& {Courtney}}]{edg03}
{Edge}, A.~C., {Ebeling}, H., {Bremer}, M., {et~al.} 2003, \mnras, 339, 913

\bibitem[{{Foley} {et~al.}(2011){Foley}, {Andersson}, {Bazin}, {de Haan},
  {Ruel}, {Ade}, {Aird}, {Armstrong}, {Ashby}, {Bautz}, {Benson}, {Bleem},
  {Bonamente}, {Brodwin}, {Carlstrom}, {Chang}, {Clocchiatti}, {Crawford},
  {Crites}, {Desai}, {Dobbs}, {Dudley}, {Fazio}, {Forman}, {Garmire}, {George},
  {Gladders}, {Gonzalez}, {Halverson}, {High}, {Holder}, {Holzapfel}, {Hoover},
  {Hrubes}, {Jones}, {Joy}, {Keisler}, {Knox}, {Lee}, {Leitch}, {Lueker},
  {Luong-Van}, {Marrone}, {McMahon}, {Mehl}, {Meyer}, {Mohr}, {Montroy},
  {Murray}, {Padin}, {Plagge}, {Pryke}, {Reichardt}, {Rest}, {Ruhl},
  {Saliwanchik}, {Saro}, {Schaffer}, {Shaw}, {Shirokoff}, {Song}, {Spieler},
  {Stalder}, {Stanford}, {Staniszewski}, {Stark}, {Story}, {Stubbs},
  {Vanderlinde}, {Vieira}, {Vikhlinin}, {Williamson}, \& {Zenteno}}]{fol11}
{Foley}, R.~J., {Andersson}, K., {Bazin}, G., {et~al.} 2011, \apj, 731, 86

\bibitem[{{Gobat} {et~al.}(2011){Gobat}, {Daddi}, {Onodera}, {Finoguenov},
  {Renzini}, {Arimoto}, {Bouwens}, {Brusa}, {Chary}, {Cimatti}, {Dickinson},
  {Kong}, \& {Mignoli}}]{gob11}
{Gobat}, R., {Daddi}, E., {Onodera}, M., {et~al.} 2011, \aap, 526, A133

\bibitem[{{Henry} {et~al.}(2010){Henry}, {Salvato}, {Finoguenov}, {Bouche},
  {Brunner}, {Burwitz}, {Buschkamp}, {Egami}, {F{\"o}rster-Schreiber},
  {Fotopoulou}, {Genzel}, {Hasinger}, {Mainieri}, {Rovilos}, \&
  {Szokoly}}]{hen10}
{Henry}, J.~P., {Salvato}, M., {Finoguenov}, A., {et~al.} 2010, \apj, 725, 615

\bibitem[{{Hurier} {et~al.}(2010){Hurier}, {Hildebrandt}, \&
  {Macias-Perez}}]{hur10}
{Hurier}, G., {Hildebrandt}, S.~R., \& {Macias-Perez}, J.~F. 2010, \aap\
  submitted, arXiv:1007.1149

\bibitem[{{Jee} {et~al.}(2009){Jee}, {Rosati}, {Ford}, {Dawson}, {Lidman},
  {Perlmutter}, {Demarco}, {Strazzullo}, {Mullis}, {B{\"o}hringer}, \&
  {Fassbender}}]{jee09}
{Jee}, M.~J., {Rosati}, P., {Ford}, H.~C., {et~al.} 2009, \apj, 704, 672

\bibitem[{{Kravtsov} {et~al.}(2006){Kravtsov}, {Vikhlinin}, \& {Nagai}}]{kra06}
{Kravtsov}, A.~V., {Vikhlinin}, A., \& {Nagai}, D. 2006, \apj, 650, 128

\bibitem[{{Marriage} {et~al.}(2011){Marriage}, {Baptiste Juin}, {Lin},
  {Marsden}, {Nolta}, {Partridge}, {Ade}, {Aguirre}, {Amiri}, {Appel},
  {Barrientos}, {Battistelli}, {Bond}, {Brown}, {Burger}, {Chervenak}, {Das},
  {Devlin}, {Dicker}, {Bertrand Doriese}, {Dunkley}, {D{\"u}nner},
  {Essinger-Hileman}, {Fisher}, {Fowler}, {Hajian}, {Halpern}, {Hasselfield},
  {Hern{\'a}ndez-Monteagudo}, {Hilton}, {Hilton}, {Hincks}, {Hlozek},
  {Huffenberger}, {Handel Hughes}, {Hughes}, {Infante}, {Irwin}, {Kaul},
  {Klein}, {Kosowsky}, {Lau}, {Limon}, {Lupton}, {Martocci}, {Mauskopf},
  {Menanteau}, {Moodley}, {Moseley}, {Netterfield}, {Niemack}, {Page},
  {Parker}, {Quintana}, {Reid}, {Sehgal}, {Sherwin}, {Sievers}, {Spergel},
  {Staggs}, {Swetz}, {Switzer}, {Thornton}, {Trac}, {Tucker}, {Warne},
  {Wilson}, {Wollack}, \& {Zhao}}]{mar11}
{Marriage}, T.~A., {Baptiste Juin}, J., {Lin}, Y., {et~al.} 2011, \apj, 731,
  100

\bibitem[{{Melin} {et~al.}(2006){Melin}, {Bartlett}, \& {Delabrouille}}]{mel06}
{Melin}, J., {Bartlett}, J.~G., \& {Delabrouille}, J. 2006, \aap, 459, 341

\bibitem[{{Mennella et al.}(2011)}]{planck2011-1.4}
{Mennella et al.} 2011, {Planck early results 03: First assessment of the Low
  Frequency Instrument in-flight performance} ({Submitted to \aap,
  [arXiv:astro-ph/1101.2038]})

\bibitem[{{Mortonson} {et~al.}(2011){Mortonson}, {Hu}, \& {Huterer}}]{mor11}
{Mortonson}, M.~J., {Hu}, W., \& {Huterer}, D. 2011, \prd, 83, 023015

\bibitem[{{Mullis} {et~al.}(2005){Mullis}, {Rosati}, {Lamer}, {B{\"o}hringer},
  {Schwope}, {Schuecker}, \& {Fassbender}}]{mul05}
{Mullis}, C.~R., {Rosati}, P., {Lamer}, G., {et~al.} 2005, \apjl, 623, L85

\bibitem[{{Piffaretti} {et~al.}(2010){Piffaretti}, {Arnaud}, {Pratt},
  {Pointecouteau}, \& {Melin}}]{pif10}
{Piffaretti}, R., {Arnaud}, M., {Pratt}, G.~W., {Pointecouteau}, E., \&
  {Melin}, J. 2010, {\aap\ submitted, arXiv:1007.1916}

\bibitem[{{Planck Collaboration}(2011{\natexlab{a}})}]{planck2011-1.1}
{Planck Collaboration}. 2011{\natexlab{a}}, {Planck early results 01: The
  Planck mission} ({Submitted to \aap, [arXiv:astro-ph/1101.2022]})

\bibitem[{{Planck Collaboration}(2011{\natexlab{b}})}]{planck2011-5.1a}
{Planck Collaboration}. 2011{\natexlab{b}}, {Planck early results 08: The
  all-sky early Sunyaev-Zeldovich cluster sample} ({Submitted to \aap,
  [arXiv:astro-ph/1101.2024]})

\bibitem[{{Planck Collaboration}(2011{\natexlab{c}})}]{planck2011-5.1b}
{Planck Collaboration}. 2011{\natexlab{c}}, {Planck early results 09:
  XMM-Newton follow-up for validation of Planck cluster candidates} ({Submitted
  to \aap, [arXiv:astro-ph/1101.2025]})

\bibitem[{{Planck HFI Core Team}(2011)}]{planck2011-1.5}
{Planck HFI Core Team}. 2011, {Planck early results 04: First assessment of the
  High Frequency Instrument in-flight performance} ({Submitted to \aap,
  [arXiv:astro-ph/1101.2039]})

\bibitem[{{Pratt} \& {Arnaud}(2002)}]{pra02}
{Pratt}, G.~W. \& {Arnaud}, M. 2002, \aap, 394, 375

\bibitem[{{Pratt} {et~al.}(2010){Pratt}, {Arnaud}, {Piffaretti},
  {B{\"o}hringer}, {Ponman}, {Croston}, {Voit}, {Borgani}, \& {Bower}}]{pra10}
{Pratt}, G.~W., {Arnaud}, M., {Piffaretti}, R., {et~al.} 2010, \aap, 511, A85

\bibitem[{Pratt {et~al.}(2007)Pratt, B{\"o}hringer, Croston, Arnaud, Borgani,
  Finoguenov, \& Temple}]{pra07}
Pratt, G.~W., B{\"o}hringer, H., Croston, J.~H., {et~al.} 2007, \aap, 461, 71

\bibitem[{{Pratt} {et~al.}(2009){Pratt}, {Croston}, {Arnaud}, \&
  {B{\"o}hringer}}]{pra09}
{Pratt}, G.~W., {Croston}, J.~H., {Arnaud}, M., \& {B{\"o}hringer}, H. 2009,
  \aap, 498, 361

\bibitem[{{Santos} {et~al.}(2010){Santos}, {Tozzi}, {Rosati}, \&
  {B{\"o}hringer}}]{san10}
{Santos}, J.~S., {Tozzi}, P., {Rosati}, P., \& {B{\"o}hringer}, H. 2010, \aap,
  521, A64

\bibitem[{{Str{\"u}der} {et~al.}(2001){Str{\"u}der}, {Briel}, {Dennerl},
  {Hartmann}, {Kendziorra}, {Meidinger}, {Pfeffermann}, {Reppin}, {Aschenbach},
  {Bornemann}, {Br{\"a}uninger}, {Burkert}, {Elender}, {Freyberg}, {Haberl},
  {Hartner}, {Heuschmann}, {Hippmann}, {Kastelic}, {Kemmer}, {Kettenring},
  {Kink}, {Krause}, {M{\"u}ller}, {Oppitz}, {Pietsch}, {Popp}, {Predehl},
  {Read}, {Stephan}, {St{\"o}tter}, {Tr{\"u}mper}, {Holl}, {Kemmer}, {Soltau},
  {St{\"o}tter}, {Weber}, {Weichert}, {von Zanthier}, {Carathanassis}, {Lutz},
  {Richter}, {Solc}, {B{\"o}ttcher}, {Kuster}, {Staubert}, {Abbey}, {Holland},
  {Turner}, {Balasini}, {Bignami}, {La Palombara}, {Villa}, {Buttler},
  {Gianini}, {Lain{\'e}}, {Lumb}, \& {Dhez}}]{str01}
{Str{\"u}der}, L., {Briel}, U., {Dennerl}, K., {et~al.} 2001, \aap, 365, L18

\bibitem[{Sunyaev \& Zeldovich(1972)}]{sun72}
Sunyaev, R.~A. \& Zeldovich, Y.~B. 1972, Comments on Astrophysics and Space
  Physics, 4, 173

\bibitem[{{Turner} {et~al.}(2001){Turner}, {Abbey}, {Arnaud}, {Balasini},
  {Barbera}, {Belsole}, {Bennie}, {Bernard}, {Bignami}, {Boer}, {Briel},
  {Butler}, {Cara}, {Chabaud}, {Cole}, {Collura}, {Conte}, {Cros}, {Denby},
  {Dhez}, {Di Coco}, {Dowson}, {Ferrando}, {Ghizzardi}, {Gianotti}, {Goodall},
  {Gretton}, {Griffiths}, {Hainaut}, {Hochedez}, {Holland}, {Jourdain},
  {Kendziorra}, {Lagostina}, {Laine}, {La Palombara}, {Lortholary}, {Lumb},
  {Marty}, {Molendi}, {Pigot}, {Poindron}, {Pounds}, {Reeves}, {Reppin},
  {Rothenflug}, {Salvetat}, {Sauvageot}, {Schmitt}, {Sembay}, {Short},
  {Spragg}, {Stephen}, {Str{\"u}der}, {Tiengo}, {Trifoglio}, {Tr{\"u}mper},
  {Vercellone}, {Vigroux}, {Villa}, {Ward}, {Whitehead}, \& {Zonca}}]{tur01}
{Turner}, M.~J.~L., {Abbey}, A., {Arnaud}, M., {et~al.} 2001, \aap, 365, L27

\bibitem[{{Vikhlinin} {et~al.}(2007){Vikhlinin}, {Burenin}, {Forman}, {Jones},
  {Hornstrup}, {Murray}, \& {Quintana}}]{vik07}
{Vikhlinin}, A., {Burenin}, R., {Forman}, W.~R., {et~al.} 2007, in Heating
  versus Cooling in Galaxies and Clusters of Galaxies, ed. {H.~B{\"o}hringer,
  G.~W.~Pratt, A.~Finoguenov, \& P.~Schuecker } ({Springer-Verlag Berlin
  Heidelberg}), 48

\bibitem[{{Williamson} {et~al.}(2011){Williamson}, {Benson}, {High},
  {Vanderlinde}, {Ade}, {Aird}, {Andersson}, {Armstrong}, {Ashby}, {Bautz},
  {Bazin}, {Bertin}, {Bleem}, {Bonamente}, {Brodwin}, {Carlstrom}, {Chang},
  {Clocchiatti}, {Crawford}, {Crites}, {de Haan}, {Desai}, {Dobbs}, {Dudley},
  {Fazio}, {Foley}, {Forman}, {Garmire}, {George}, {Gladders}, {Gonzalez},
  {Halverson}, {Holder}, {Holzapfel}, {Hoover}, {Hrubes}, {Jones}, {Joy},
  {Keisler}, {Knox}, {Lee}, {Leitch}, {Lueker}, {Luong-Van}, {Marrone},
  {McMahon}, {Mehl}, {Meyer}, {Mohr}, {Montroy}, {Murray}, {Padin}, {Plagge},
  {Pryke}, {Reichardt}, {Rest}, {Ruel}, {Ruhl}, {Saliwanchik}, {Saro},
  {Schaffer}, {Shaw}, {Shirokoff}, {Song}, {Spieler}, {Stalder}, {Stanford},
  {Staniszewski}, {Stark}, {Story}, {Stubbs}, {Vieira}, {Vikhlinin}, \&
  {Zenteno}}]{wil11}
{Williamson}, R., {Benson}, B.~A., {High}, F.~W., {et~al.} 2011, \apj\
  submitted, ArXiv:1101.1290

\end{thebibliography}

\raggedright
\end{document}